\documentclass[floatfix,prx,twocolumn,letterpaper,lengthcheck,superscriptaddress,showpacs,amssymb,amsmath,amsfonts,aps,altaffilletter,nofootinbib,nopreprintnumbers,showpacs,longbibliography]{revtex4-1}

\usepackage{color}
\usepackage{hyperref}
\usepackage{amsmath}
\usepackage{amsthm}
\usepackage{multirow}
\usepackage{graphicx}
\usepackage{caption}
\usepackage{subcaption}
\usepackage{placeins}

\usepackage{graphicx}
\usepackage{calc}
\usepackage{tabularx}
\usepackage{amsmath}
\usepackage{amssymb}
\usepackage{hyperref} 
\usepackage{float} 

\usepackage{mathrsfs}
\DeclareSymbolFontAlphabet{\mathrsfs}{rsfs}
\newcommand{\scrM}{\mathrsfs{M}}

\newcommand{\Surf}{\ensuremath{\mathcal{S}}}
\def\insubscript{\rm inner}
\def\outsubscript{\rm outer}

\newcommand{\Sout}{\Surf_{\outsubscript}}

\newcommand{\Sin}{\Surf_{\insubscript}}

\newcommand{\Sone}{\Surf_{1}}

\newcommand{\Stwo}{\Surf_{2}}

\newcommand{\HH}{\ensuremath{\mathcal{H}}}
\newcommand{\Hout}{\HH_{\outsubscript}}
\newcommand{\Hin}{\HH_{\insubscript}}
\newcommand{\Hone}{\HH_{1}}
\newcommand{\Htwo}{\HH_{2}}

\def\twoq{\widetilde{q}}
\def\twoepsilon{\widetilde{\epsilon}}

\def\tname{T}
\def\ttouch{\tname_{\rm touch}}
\def\tbifurcate{\tname_{\rm bifurcate}}

\def\MM{\mathcal{M}}
\def\MADM{M_\text{ADM}}

\newcommand{\Mirr}{M_{\rm irr}}

\newtheorem{defn}{Definition}

\begin{document}

\title[]{Horizons in a binary black hole merger II: Fluxes, multipole
  moments and stability}

\author{Daniel Pook-Kolb} 
\affiliation{Max-Planck-Institut f\"ur Gravitationsphysik (Albert
  Einstein Institute), Callinstr. 38, 30167 Hannover, Germany}
\affiliation{Leibniz Universit\"at Hannover, 30167 Hannover, Germany}

\author{Ofek Birnholtz}
\affiliation{Department of Physics, Bar-Ilan University, Ramat-Gan
  5290002, Israel}

\author{Jos\'e Luis Jaramillo}
\affiliation{Institut de Math\'ematiques de Bourgogne (IMB), UMR 5584,
  CNRS,Universit\'e de Bourgogne Franche-Comt\'e, F-21000 Dijon,
  France}

\author{Badri Krishnan} 
\affiliation{Max-Planck-Institut f\"ur Gravitationsphysik (Albert
  Einstein Institute), Callinstr. 38, 30167 Hannover, Germany}
\affiliation{Leibniz Universit\"at Hannover, 30167 Hannover, Germany}

\author{Erik Schnetter}
\affiliation{Perimeter Institute for Theoretical Physics, Waterloo, 
  ON N2L 2Y5, Canada}
\affiliation{Physics \& Astronomy Department, University of Waterloo,
  Waterloo, ON N2L 3G1, Canada}
\affiliation{Center for Computation \& Technology, Louisiana State
  University, Baton Rouge, LA 70803, USA}


\begin{abstract}

  We study in detail the dynamics and stability of marginally trapped
  surfaces during a binary black hole merger.  This is the second in a
  two-part study.  The first part studied the basic geometric aspects
  of the world tubes traced out by the marginal surfaces and the
  status of the area increase law.  Here we continue and study the
  dynamics of the horizons during the merger, again for the head-on
  collision of two non-spinning black holes.  In particular we follow
  the spectrum of the stability operator during the course of the
  merger for all the horizons present in the problem and implement
  systematic spectrum statistics for its analysis.  We also study more
  physical aspects of the merger, namely the fluxes of energy which
  cross the horizon and cause the area to change.  We construct a
  natural coordinate system on the horizon and decompose the various
  fields appearing in the flux, primarily the shear of the outgoing
  null normal, in spin weighted spherical harmonics.  For each of the
  modes we extract the decay rates as the final black hole approaches
  equilibrium.  The late part of the decay is consistent with the
  expected quasi-normal mode frequencies, while the early part
  displays a much steeper fall-off.  Similarly, we calculate the decay
  of the horizon multipole moments, again finding two different
  regimes.  Finally, seeking an explanation for this behavior,
  motivated by the membrane paradigm interpretation, we attempt to
  identify the different dynamical timescales of the area increase.
  This leads to the definition of a ``slowness parameter'' for
  predicting the onset of transition from a faster to a slower decay.
 
\end{abstract}

\maketitle

\section{Introduction}
\label{sec:intro}

In classical general relativity, black holes are perfect absorbers.
They grow inexorably by absorbing matter and/or radiation from their
surroundings. Emission of electromagnetic or gravitational radiation
occurs due to interactions of the black hole with surrounding
spacetime or matter.  Gravitational waves are emitted due to
non-stationarities and non-linearities of the spacetime metric in the
region around the black hole.  Black holes have an additional special
feature which does not hold for other physical objects, namely a very
special set of equilibrium states determined by only two parameters in
astrophysical contexts. In other words, astrophysical black holes
within standard general relativity have no hair.  Normal physical
objects reach equilibrium by both absorbing and emitting, but black
holes do not have that luxury.  Not only must they only absorb, but
they must absorb very selectively so that the absorbed radiation
precisely cancels any hair it might initially have.

This picture applies to a binary black hole merger.  When the final
remnant black hole is initially formed, its horizon is highly
distorted but its final state is that of a simple Kerr black
hole. This process of reaching equilibrium from its initial state at
formation must follow the process of selective absorption mentioned
above.  This process of reaching equilibrium is often referred to as
the black hole ``radiating away its hair''. This is accurate when one
considers a sufficiently large spacetime region containing the black
hole; after all, it is not just the horizon that reaches equilibrium,
but rather the spacetime itself in a neighborhood of the horizon.
However, ``radiating away hair'' is not an apt description for the
horizon itself in classical general relativity.

The issue of how a black hole knows precisely how much radiation to
absorb at any given time, is an important one in general relativity.
From a mathematical perspective, it touches on the question of the
stability of the Kerr black hole in full non-linear general
relativity.  From a theoretical physics viewpoint, any deviations of
the final state from Kerr might indicate support for alternate
theories of gravity. As we have argued in the previous paragraph, this
issue of the final state is intimately connected with the in-falling
energy flux through the horizon.  One important goal of analytic or
theoretical studies is thus to discover universalities in the approach
to equilibrium of a black hole horizon in full non-linear general
relativity.  These universalities might be reflected in the rates of
exponential or power-law decay. Gravitational wave observations of
binary black hole mergers offer opportunities for testing these
predictions observationally.

A useful way of approaching these problems is via the study of
marginally trapped surfaces.  These are special spherical surfaces for
which outgoing light rays have vanishing convergence.  These surfaces
are well suited for describing not only stationary black holes, but
also binary mergers and other dynamical processes involving black
holes. The entire process of merger and approach to equilibrium can be
understood in terms of marginally trapped surfaces.  Recent numerical
studies have discovered new geometric and topological features of
marginally trapped surfaces in binary black hole mergers.  These
include their behavior under time evolution, the status of the area
increase law, and the presence of topological features such as cusps
and knots.  These numerical results rely on a new method for locating
marginally outer trapped surfaces
\cite{Pook-Kolb:2018igu,Pook-Kolb:2018igu,PhysRevD.100.084044}, and
the physical results are based on the formalism of quasi-local
horizons.  This formalism is based on the world tube of marginally
trapped surfaces and it provides a coherent way of studying various
aspects of black hole physics quasi-locally
\cite{Booth:2005qc,Ashtekar:2004cn,Gourgoulhon:2005ng,Jaramillo:2011zw,Faraoni:2015pmn,Visser:2009xp,Hayward:2000ca}.
For our purposes, it is important that there exist exact flux formulae
for these horizons within full general relativity, which quantify the
amount of energy and radiation crossing the horizon, and relate it to
the change in horizon area \cite{Ashtekar:2002ag,Ashtekar:2003hk}.
The flux due to gravitational radiation is positive definite and
always causes the area to increase.  This is analogous to the well
known Bondi mass-loss formula at null infinity in the Bondi-Sachs
framework describing the energy carried away by gravitational
radiation.

It turns out that in these astrophysical situations, the fluxes
falling through the horizon are highly correlated with the fluxes at
infinity which can be observed by gravitational wave detectors
\cite{Jaramillo:2011rf,Jaramillo:2012rr,Rezzolla:2010df,Gupta:2018znn,Prasad:2020xgr}. This
might appear surprising at first glance since the horizons are
causally disconnected from observers outside the event horizon.
However, in these astrophysical situations the source of the
in-falling radiation and the outgoing radiation are one and the same,
namely non-linearities and non-stationarities in the spacetime region
near (but outside) the black holes.  Thus, a better understanding of
the horizon fluxes might help us to quantify these correlations
better.  Eventually, one might be able to observationally infer
properties of spacetime regions hidden behind event horizons.

The goal of this paper is to study, via numerical simulations, horizon
fluxes in binary black hole mergers, and the approach to equilibrium.
The basic scenario outlining how marginally trapped surfaces merge has
been established in
\cite{Pook-Kolb:2018igu,Pook-Kolb:2018igu,PhysRevD.100.084044}. The
present series of papers follows up on these results by studying
physical and geometrical properties of marginally trapped surfaces and
their time evolution.  The first paper (henceforth paper I) has
studied basic properties of these world tubes including their
signature and the status of the area increase law. The goal here is to
study in detail physical aspects of these world tubes. These include
energy fluxes across the world tubes, their decay rates as the final
black hole approaches equilibrium, the evolution of the horizon
multipole moments, and their stability properties.  While we often
refer to paper I (and the reader might benefit by having a copy of
that paper at hand), this paper is meant to be mostly self-contained.

The plan for the rest of this paper is as follows.
Sec.~\ref{sec:quasilocal} sets up notation and briefly summarizes some
of the basic notions and results that we shall use later. Paper I has
already summarized the main definitions and concepts of quasi-local
horizons that we employ.  Here we shall summarize results pertaining
to the horizon fluxes, the stability operator and the multipole
moments. Especially important will be the construction of an invariant
coordinate system on the horizon which will be used to decompose
various fields on the horizon.  Sec.~\ref{sec:stability} discusses the
stability of the various MOTSs.  The stability here refers to the
properties of a MOTS under small outward deformations, and is governed
by an elliptic operator.  The horizon will be stable if this operator
is invertible, i.e. when its spectrum does not contain zero.  This
leads us then to analyze the spectral properties of the operator,
yielding what might be called the stability spectrum of the MOTS and
pushing forward the study of the full MOTS-spectral problem formulated
in \cite{Jaramillo:2013rda,Jaramillo:2014oha,Jaramillo:2015twa} in
particular introducing a discussion in terms of spectrum statistics.

Sec.~\ref{sec:flux} addresses the question of why the area changes,
namely due to the flux of gravitational radiation across the
horizon. The most important part of the radiation flux is the shear
which, just like the gravitational radiation observed by gravitational
wave detectors, is a symmetric tracefree tensor, except that it lives
on the horizon.  The horizon, being a non-null surface, also has
another contribution to the flux from a vector field on the horizon.
We study the multipolar decomposition of both of these contributions.
We then connect the decay rate of the flux to the quasi-normal mode
frequencies associated with the final black hole.
Sec.~\ref{sec:moments} presents the evolution of the horizon multipole
moments.  The multipole moments capture the deviation of the horizon
from a simple Schwarzschild geometry (or Kerr, if the black holes had
been rotating).  Thus, the evolution of the multipole moments in time
tells us about how the two individual black holes become increasingly
distorted, and how the final black hole approaches equilibrium.  This
is, of course closely connected with the fluxes discussed in
Sec.~\ref{sec:flux}.  Sec.~\ref{sec:slowness} offers a tentative
explanation for why we have two regimes in the approach to
equilibrium.  It shows that the non-linear effects dominate in the
steep decay regime at early times, while the later time is consistent
with linear behavior.  Sec.~\ref{sec:conclusions} concludes by
discussing open questions and possible directions for future work. The
mathematical issues discussed in Sec.~\ref{sec:stability} (namely
spectral theory) are quite different from the topics of
Secs.~\ref{sec:flux} and \ref{sec:moments} (fluxes, multipole moments,
quasi-normal modes, and non-linearities); they can thus be read quite
independently of each other.

\section{Basic Notions}
\label{sec:quasilocal} 

\subsection{Marginally trapped surfaces and dynamical horizons}

The basic notions of marginally trapped surfaces and dynamical
horizons were already summarized in paper I.  Several review articles
on the subject are also available
\cite{Booth:2005qc,Ashtekar:2004cn,Gourgoulhon:2005ng,Faraoni:2015pmn,Visser:2009xp,Hayward:2000ca}.
We shall therefore be very brief with the basic definitions.  The
focus will be on the flux laws, multipole moments and the stability
operator.

Let spacetime be modeled as a 4-dimensional manifold $\scrM$
equipped with a Lorentzian metric $g_{ab}$ with signature $(-,+,+,+)$.
We shall only consider vacuum spacetimes.  Let $\nabla_a$ be the
derivative operator compatible with $g_{ab}$.  Let $\Surf$ be a closed
2-dimensional spacelike manifold immersed in $\scrM$. $\Surf$ is
taken to be orientable and of spherical topology. Let $\twoq_{ab}$,
$\twoepsilon$, and $\mathcal{D}_a$ be the intrinsic Riemannian metric
on $\Surf$, the volume 2-form, and the corresponding derivative
operator, respectively.  The intrinsic scalar curvature of $\Surf$ will
be denoted $\mathcal{R}$, its area $A_\Surf$, and the Laplacian on
$\Surf$ is $\Delta_\Surf$.

The outgoing and ingoing future directed null-normals to $\Surf$ will
be denoted by $\ell^a$ and $n^a$ respectively.  We will tie the
normalizations of the null normals together by requiring
$\ell\cdot n = -1$. Finally, given a complex null vector $m^a$ tangent
to $\Surf$ satisfying $m\cdot\bar{m}=1$, we obtain a null-tetrad
$(\ell,n,m,\bar{m})$.

The expansions $\Theta_{(\ell)}$ and $\Theta_{(n)}$ of $\ell^a$ and
$n^a$ are respectively
\begin{equation}
  \Theta_{(\ell)} = \twoq^{ab}\nabla_a\ell_b\,,\quad \Theta_{(n)} = \twoq^{ab}\nabla_an_b\,.  
\end{equation}
The shears $\sigma_{(\ell)}$ and $\sigma_{(n)}$ of $\ell^a$ and $n^a$,
respectively, are
\begin{equation}
  \sigma_{(\ell)} = m^am^b\nabla_a\ell_b\,,\quad \sigma_{(n)} = m^am^b\nabla_an_b\,. 
\end{equation}
We shall usually not need $\sigma_{(n)}$ in this paper, and thus we
shall often refer to $\sigma_{(\ell)}$ just as the shear $\sigma$.

The other important field is the connection 1-form on the normal
bundle of $\Surf$:
\begin{equation}
 \omega_a = -n_bq_{a}^c\nabla_c\ell^b\,. 
\end{equation}
It can be shown that $\omega_a$ relates to the angular momentum
associated with $\Surf$ (see
e.g. \cite{Ashtekar:2001is,Ashtekar:2003hk}).  In this paper we
consider only non-spinning black holes. Thus while we will
occasionally mention $\omega_a$ where appropriate, all of our results
have $\omega_a=0$.

$\Surf$ is said to be a \emph{future-marginally-outer-trapped} surface
if $\Theta_{(\ell)}=0$ and $\Theta_{(n)}<0$.  If $\Theta_{(n)}>0$,
then $\Surf$ is said to be \emph{past-marginally-outer-trapped}.  A
surface satisfying only $\Theta_{(\ell)}=0$ with no restriction on
$\Theta_{(n)}$ is called a marginally outer trapped surface, or MOTS in
short.

It is clear that a MOTS is a geometric concept in a spacetime, and
makes no reference to any spacelike Cauchy surfaces or time
coordinate.  Nevertheless, one can think of a Cauchy surface as a
convenient means of locating a MOTS: They can be located on a
spacelike Cauchy surface $\Sigma$ equipped with a 3-metric and
extrinsic curvature, and well known numerical methods exist for this.
The canonical choice of null normals for $\Surf$ immersed in $\Sigma$
is
\begin{equation}
  \label{eq:normals}
  \ell^a = \frac{1}{\sqrt{2}}\left( T^a + R^a\right)\,,\quad
  n^a = \frac{1}{\sqrt{2}}\left( T^a - R^a\right)\,.
\end{equation}
Here $R^a$ is the unit spacelike normal to $\Surf$ (and tangent to
$\Sigma$), while $T^a$ is the unit timelike normal to $\Sigma$.  We
use a numerical method recently developed in
\cite{Pook-Kolb:2018igu,PhysRevD.100.084044}, capable of locating
highly distorted surfaces; our implementation is available at
\cite{pook_kolb_daniel_2019_2591105}.  This method is an extension of
the widely used method developed in
\cite{Thornburg:2003sf,Thornburg:2006zb,Thornburg:1995cp,Thornburg:1995cp,Shoemaker:2000ye,Lin:2007cd}.
Our numerical calculation use \emph{Einstein Toolkit}
\cite{Loffler:2011ay, EinsteinToolkit:web}. We use \emph{TwoPunctures}
\cite{Ansorg:2004ds} to set up initial conditions and an axisymmetric
version of \emph{McLachlan} \cite{Brown:2008sb} to solve the Einstein
equations, which uses \emph{Kranc} \cite{Husa:2004ip, Kranc:web} to
generate efficient C++ code.
Results in this paper are obtained from simulations with spatial
resolutions $1/\Delta x = 480$ running until
$\tname_\text{max} = 20\,\MM$ and $1/\Delta x = 60$ running until
$\tname_\text{max} = 50\,\MM$, where $\MM := \MADM/1.3$ is our
simulation time unit.  For brevity, we will occasionally state
simulation times using lowercase $t := T/\MM$.  Here $\MM$ is a
suitable mass scale in the problem.
Further details of the simulation
specific to our problem are detailed in \cite{PhysRevD.100.084044}.

The initial configuration is the same as that used in paper I and in
\cite{PhysRevD.100.084044}.  We use the Brill-Lindquist construction
\cite{Brill:1963yv}, i.e. the initial data is conformally flat and
time symmetric. The initial data has two non-spinning black holes with
vanishing linear momentum.  The ``bare masses'' are $m_1=0.5$ and
$m_2=0.8$ with the total ADM mass being $\MADM=1.3$. The initial
separation $d_0$ is $d_0/\MADM = 1$.  At the initial time, there are
two disjoint horizons $\Sone$ and $\Stwo$ with $\Stwo$ being the
larger one.  The common horizon forms at a time $\tbifurcate$ shortly
after the simulation starts and splits into inner and outer surfaces,
$\Sin$ and $\Sout$, respectively.  The world tubes of these horizons are
shown in Fig.~1 of paper I.

The 3-dimensional world tube traced out by the MOTSs is taken as a
bonafide geometric object in its own right and we attempt to
understand its physical and geometric properties.  The pioneering work
by Hayward \cite{Hayward:1993wb} was an important step in this
direction. Another important aspect is a detailed study of the case
when the world tube is null, i.e. just like the stationary
Schwarzschild and Kerr solutions, the black hole is not absorbing
matter/energy and not increasing in area.  This can be viewed as an
approximation in suitable physical situations (an excellent
approximation in many cases), or as the limiting case asymptotically
as the black hole reaches equilibrium.  The basic definition of a
non-expanding horizon and its extensions to an isolated horizon has
been summarized in paper I.  A detailed understanding of this case has
been achieved and an extensive literature on isolated horizons is
available (see
e.g. \cite{Ashtekar:1998sp,Ashtekar:1999yj,Ashtekar:2001is,Ashtekar:2001jb,Lewandowski:1999zs,Lewandowski:2018khe,Korzynski:2004gr,Ashtekar:2000hw,Krishnan:2012bt,Booth:2001gx,Booth:2012xm}).
For the dynamical case, we need to consider a general world tube of
arbitrary signature which will be called a dynamical horizon.
Additional qualifiers such as timelike or spacelike, and future and
past (depending on the sign of $\Theta_{(n)}$) will be included as
required.

\subsection{Variations and the stability operator}
\label{subsec:stability}

Given a MOTS $\Surf$ on a Cauchy surface $\Sigma$ and a choice of
lapse and shift, i.e. a time evolution vector, consider the behavior
of the MOTS under time evolution.  If the MOTS were to evolve smoothly
under this time evolution, it would trace out a smooth 3-dimensional
world tube.  In the well known stationary solutions, e.g. the
Schwarzschild or Kerr black holes, the event horizons are foliated by
MOTSs.  If the world tube does exist also in fully dynamical
situations, then it is possible to formulate black hole physics and
thermodynamics in various physical scenarios.  Seminal work by Hayward
in 1994 introduced the notion of trapping horizons
\cite{Hayward:1993wb} and showed how one could formulate the laws of
black hole thermodynamics in this framework for dynamical black holes.
Similarly, horizon fluxes were studied in
\cite{Ashtekar:2002ag,Ashtekar:2003hk} and shown to be manifestly
positive definite.  In this early work on this topic, it was usually
assumed that this smooth world tube exists in full non-linear general
relativity.  This was a reasonable assumption, especially given the
fact that MOTSs were already widely used in numerical relativity for
locating and extracting physical black hole parameters
\cite{Dreyer:2002mx}.  In these numerical simulations the apparent
horizons were generally found to evolve smoothly.  The mathematical
conditions under which a MOTS evolves smoothly were found in 2005
\cite{Andersson:2005gq,Andersson:2007fh,Booth:2006bn}. A central role
in these proofs is played by the stability operators associated with a
MOTS and their eigenvalues, which we now describe.

The starting point here is the notion of the variation of a MOTS
\cite{Newman:1987}.  One chooses a vector field $X^a$ along which
$\Surf$ is to be varied, thereby obtaining a family of surfaces
$\Surf_\lambda$ at least for small values of $\lambda$. Starting with
a point $p$ on $\Surf$, varying $\lambda$ yields a curve with $X^a$ as
the tangent vector at $p$; $\Surf_{\lambda=0}$ is identified with
$\Surf$ itself.  Variations tangent to $\Surf$ do not play an
important role here, and we take $X^a$ to be orthogonal to
$\Surf$. Given this family $\Surf_\lambda$ depending smoothly on
$\lambda$, one can consider \emph{variations} of geometric quantities
on $\Surf$.  For a MOTS, the quantity of interest is the expansion
$\Theta_{(\ell)}$. For each $\Surf_\lambda$, we define null normals
just as for $\Surf$ itself. The expansion can be computed for each
value of $\lambda$ and then differentiated.  This defines the
variation of $\Theta_{(\ell)}$ along $X^a$, which is denoted
$\delta_X\Theta_{(\ell)}$. This is not be confused with usual
derivatives of $\Theta_{(\ell)}$. In particular,
$\delta_{\psi X}\Theta_{(\ell)} \neq \psi\delta_X\Theta_{(\ell)}$ when
$\psi$ is not a constant.  This leads to the definition of the
stability operator $L$ acting on functions
$\psi:\Surf\rightarrow \mathbb{R}$ as
\begin{equation}
  L^{(X)}[\psi] := \delta_{\psi X}\Theta_{(\ell)}\,.
\end{equation}
Since $X^a$ is orthogonal to $\Surf$, given a choice of the null
normals $(\ell^a,n^a)$, we can write
\begin{equation}
  X^a = b\ell^a + cn^a \,,
\end{equation}
where $b$ and $c$ are functions on $\Surf$.  We see then that there is
not just a single stability operator, but several depending on the
normal direction.  This is why we label the stability operator
$L^{(X)}$ with $X$.

One case is well known and easy to understand, namely when $X^a$ is
along $\ell^a$.  This should just be the Raychaudhuri equation, and
indeed, setting $\Theta_{(\ell)}=0$ and assuming spacetime to be
vacuum leads to
\begin{equation}
  L^{(\ell)}[\psi] = \delta_{\psi\ell}\Theta_{(\ell)} = -2\left|\sigma\right|^2\psi\,.
\end{equation}
Clearly, if $\psi$ is positive, then this variation will be
negative. Moreover, this variation is linear in $\psi$ and does not
involve any derivatives.  The other component of the variation is
along $n^a$; it will be convenient to consider the outgoing direction
$-n^a$ instead.  This turns out to lead to a second order elliptic
operator:
\begin{eqnarray}
  L^{(-n)}[\psi]  &=& \left(-\Delta_{\Surf} + 2\omega^a\mathcal{D}_a\right)\psi \nonumber \\
  &+& \left( \frac{1}{2}\mathcal{R} + \mathcal{D}_a\omega^a - \omega^a\omega_a\right) \psi\,.
\end{eqnarray}
The presence of the first derivative causes this operator to be
non-self-adjoint.  We will have $\omega=0$ in this paper, whence this
simplifies to a self-adjoint operator
\begin{equation}\label{eq:Ln_selfadjoint}
  L^{(-n)}[\psi]  = \left(-\Delta_{\Surf} + \frac{1}{2}\mathcal{R} \right)\psi\,.
\end{equation}
We have seen that the variation along $\ell^a$ is ``negative''. On the
other hand, since $-\Delta_\Surf$ has positive eigenvalues, the
variation along $-n^a$ is seen to be positive if $\mathcal{R}$ is
positive (this shall not always be the case in this paper).

In numerical simulations, MOTSs are found on Cauchy surfaces in the
course of a time evolution.  Thus, if $\Surf$ lies on a spacelike
Cauchy surface $\Sigma$, and if $R^a$ is the unit outgoing spacelike
vector normal to $\Surf$, then it is natural to look at variations
along $R^a$. This leads to the stability operator associated with
$\Sigma$:
\begin{equation}
  L_\Sigma[\psi] := \sqrt{2}\;\delta_{\psi R}\Theta_{(\ell)}\,,
\end{equation}
where we used the freedom to choose a factor of $\sqrt{2}$ to simplify
the following expressions.
We label this stability operator by $\Sigma$ instead of
$L^{(\sqrt{2}R)}$ to emphasize the connection with the Cauchy surface.
Since $R^a = (\ell^a-n^a)/\sqrt{2}$, we have (setting $\omega_a=0$)
\begin{equation}\label{eq:Ln_sigma}
  L_\Sigma[\psi] = \delta_{\psi \ell^a - \psi n^a}\Theta_{(\ell)}
        = \left(-\Delta_{\Surf} + \frac{1}{2}\mathcal{R} - 2\left|\sigma\right|^2\right)\psi\,.
\end{equation}
Since $L_\Sigma$ and $L^{(-n)}$ are elliptic operators on a compact
manifold, they have a discrete spectrum.  In general these spectra are
complex (due to the first derivative term involving
$\omega_a$). However the eigenvalue with smallest real part can be
shown to be real, and is known as the principal eigenvalue
$\Lambda_0$.  The corresponding eigenfunction $\phi_0$ can be chosen
to be positive.  We note that the eigenvalues do not depend on the
scaling of the null normals.  If the null-normals are rescaled
according to $\ell \rightarrow f\ell$, $n\rightarrow f^{-1}n$, then
$L^{(-n)}$ undergoes a similarity transformation:
$L^{(-n)} \rightarrow fL^{(-n)}f^{-1}$.  The eigenfunctions of
$L^{(-n)}$ are scaled by $f$ but its eigenvalues are unaffected.

We now summarize some results and their connection to properties of
the various horizons that we have already encountered in paper I.
First we need a definition.
\begin{defn}[Strictly-Stably-Outermost]
  A MOTS $\Surf$ is said to be strictly-stably-outermost along a
  direction $X^a$ normal to $\Surf$ if there exists some $\psi\geq 0$
  such that $\delta_{\psi X}\Theta_{(\ell)} \geq 0$, and
  $\delta_{\psi X}\Theta_{(\ell)}$ does not vanish everywhere.
\end{defn}
This turns out to be equivalent to the principal eigenvalue being
positive definite: $\Lambda_0>0$.  If $\Lambda_0>0$ then we can choose
$\psi$ to be the lowest eigenfunction, and the condition
$\delta_{\psi X}\Theta_{(\ell)} > 0$ follows.  The converse is shown
in \cite{Andersson:2007fh}.  The principal eigenvalue itself depends
on the direction of $X^a$: it is largest for $X^a=-n^a$, and decreases
as $X^a$ turns towards $\ell^a$.  Two results are important for our
purposes:
\begin{itemize}
\item \emph{Starting with a MOTS on $\Sigma$, it evolves smoothly in
    time as long as $L_\Sigma$ is invertible, i.e. none of its
    eigenvalues vanish.  As a special case, this holds if
    $\Lambda_0>0$ whence all other eigenvalues also have positive real
    parts.}
\end{itemize}
The signature is also restricted if $\Lambda_0>0$:
\begin{itemize}
\item \emph{Let $\Surf$ be a strictly-stably-outermost MOTS. The world
    tube, i.e. the dynamical horizon, generated by the time evolution
    of $\Surf$ is spacelike if $|\sigma|^2$ is non-zero somewhere on
    $\Surf$.}
\end{itemize}
In our simulation, this scenario applies for the individual dynamical
horizons and for the outer common horizon.  All of these turn out to
be strictly-stably-outermost and, as we saw in paper I, they are all
spacelike.  The inner horizon is, as in other aspects, much more
interesting. It has $\Lambda_0<0$, and as we saw in paper I, its
signature is not restricted to be spacelike.  The spectra of
$L_\Sigma$ and $L^{(-n)}$ will be described in detail in
Sec.~\ref{sec:stability}.

\subsection{Invariant coordinates on an axisymmetric horizon}
\label{subsec:coordinates}

For physical applications to be studied below, it will be important to
decompose various fields on the horizons which have topology
$S^2\times\mathbb{R}$.  These fields will be scalar, vector and second
rank tensors.  For a given MOTS $\Surf$, some important geometric
fields of interest are the intrinsic curvature scalar $\mathcal{R}$,
the rotational 1-form $\omega_a$ and the shear.  Thus, it is very
important to have a canonical notion of scalar, vector and tensor
spherical harmonics or equivalently, spin weighted spherical
harmonics.  Different choices of spherical coordinates $(\theta,\phi)$
on a MOTS will in general yield different multipolar decompositions.
On an axisymmetric horizon, it turns out to be possible to construct
an invariant coordinate system following \cite{Ashtekar:2004gp}.

We exploit the manifest axisymmetry present in our calculations,
i.e. the existence of an axial vector $\varphi^a$ which preserves the
2-metric $q_{ab}$ on the horizon.  For an axisymmetric surface $\Surf$
of spherical topology $\Surf$ with area $A_\Surf$ and radius
$R_\Surf = \sqrt{A_\Surf/4\pi}$, we construct a coordinate system
$(\theta,\phi)$ adapted to $\varphi^a$.  We assume that $\varphi^a$
vanishes at precisely 2 points (the poles), and has closed integral
curves. The coordinate $\phi$ is the affine parameter along $\phi^a$,
taken to be in the range $[0,2\pi)$; we still need to fix the points
with $\phi=0$, which we shall do shortly.  Second, the analog of
$\cos\theta$ is a coordinate $\zeta$ defined as follows:
\begin{equation}
  \label{eq:zeta-defn}
  \mathcal{D}_a\zeta = \frac{4\pi}{A_\Surf}\widetilde{\epsilon}_{ba}\varphi^b\,,\quad \oint_\Surf\zeta\,dA = 0\,.
\end{equation}
It follows obviously that $\mathcal{D}^a\zeta$ is orthogonal to
$\varphi^a$ and its integral curves are the lines of longitude
connecting the two poles.  Fix any one of these curves, and set
$\phi=0$ on it; this specifies $\phi$ completely.  It is then
straightforward to show that the 2-metric on $\Surf$ can be written as
\begin{equation}
\label{eq:canonical-metric}
  ds^2_q = R_\Surf^2\left(\frac{d\zeta^2}{F} + Fd\phi^2 \right) \,,
\end{equation}
where
\begin{equation}
  F(\zeta) = \frac{4\pi\varphi_a\varphi^a}{A_\Surf}\,,
\end{equation}
and it can be shown that $-1<\zeta<1$ so that we can set
$\cos\theta=\zeta$.

We can now write the spin weighted spherical harmonics in terms of
$(\theta,\phi)$.  It is important to note that the orthogonality
relationships between the spherical harmonics continue to hold with
the natural volume element on $\Surf$: in the volume element for the
metric in Eq.~(\ref{eq:canonical-metric}), the factors of $F$ cancel
out.  Thus, the volume element is identical to that of a fictitious
canonical round 2-sphere metric
\begin{equation}
  q_{ab}^{(0)} = R_\Surf^2\left(d\theta^2 + \sin^2\theta d\phi^2\right)\,.
\end{equation}
Spherical harmonics, including the spin weighted spherical harmonics,
can be constructed in the usual way, but now using this canonical
metric.  Finally, a natural choice for the null vector $m$ is
\begin{equation}
  \label{eq:m}
  m = \frac{R_\Surf}{\sqrt{2}}\left(\frac{d\zeta}{\sqrt{F}} + i\sqrt{F}d\phi\right) \,.
\end{equation}
Thus, we have a complete null tetrad where $(\ell,n)$ is given by
Eq.~(\ref{eq:normals}) and $m$ is given here.

Having constructed the preferred coordinates on a given MOTS, let us
now look at its time evolution and let $\HH$ be the dynamical
horizon. For most of our results, the invariant coordinates described
above suffice: at each instant of time, we can locate the
axisymmetric MOTS, construct the invariant coordinate system,
calculate the relevant physical quantity in this coordinate system,
and then consider it as a function of time. There is no need to
explicitly consider the problem of identifying points at different
instants of time.  In future work, when we do not have axisymmetry,
this issue will be especially important if we wish to have a canonical
notion of time evolution on $\HH$.  Even in this paper, it will be
useful to clarify what one means by time evolution on $\HH$.

Let us label the MOTSs on $\HH$ by a parameter $\lambda$ (which in our
case can just be the time coordinate of the numerical evolution) and
let us consider a vector field $X^a$ tangent to $\HH$.  In principle
it need not necessarily be orthogonal to the MOTSs. The role of $X^a$
is to evolve geometric fields from one MOTS to the next.  In order to
talk about ``time evolution'' of fields and multipole moments on a
dynamical horizon, it is necessary to have a canonical choice of
$X^a$. One obvious choice is to take $X^a$ such that it preserves
the foliation of $\HH$ by MOTSs, and is orthogonal to the MOTSs.
We shall call this vector field $V^a$.  For
concreteness, take $\HH$ to be spacelike everywhere so that we have a
unit spacelike normal $\widehat{r}^a$ to each MOTS.  Then,
orthogonality of $V^a$ to the MOTSs implies
\begin{equation}
  V^a = a \widehat{r}^a
\end{equation}
with $a$ being a function on $\HH$.  $V^a$ preserves the foliation if
we can choose $V^a\partial_a\lambda = 1$, and this naturally restricts
$a$.  We also require $V^a$ to preserve the axial symmetry
$\varphi^a$: $\mathcal{L}_\varphi V^a = 0$.

There are many situations where the above choice of $V^a$ as evolution
vector is not appropriate, and we need to add a shift vector $N^a$
tangent to $\Surf$:
\begin{equation}
  X^a = a\widehat{r}^a + N^a\,.
\end{equation}
An obvious example is when we have spinning black holes, so that we
might need to add an angular velocity term:
$X^a = a\widehat{r}^a + \Omega\varphi^a$.  Even for non-spinning black
holes, it might be natural to have a non-vanishing shift vector.  A
general construction for $X^a$ satisfying certain natural conditions
is given in \cite{Ashtekar:2013qta} to determine the ``lapse'' and
``shift'' for $X^a$ as we move from one MOTS to the next.  Let us
briefly summarize the construction, specializing only later to the
case when each MOTS is axisymmetric. An important condition, it turns
out, is to choose $X^a$ such that it preserves divergence free vector
fields.  The MOTSs are changing in area and thus the volume 2-form
$\twoepsilon_{ab}$ is varying in time. We can think of this variation
as being composed of i) an overall, homogeneous change corresponding
to the overall area change, and ii) inhomogeneous variations on
smaller scales which average away to zero on each MOTS.  It turns out
that the right condition is to choose $X^a$ such that
\begin{equation}
  \mathcal{L}_X\left( \frac{\twoepsilon_{ab}}{A_\Surf}\right)  = 0\,.
\end{equation}
Note that the quantity $\twoepsilon_{ab}/A_\Surf$ integrates to unity
and contains the local inhomogeneous fluctuations in the area element
on $\Surf$. Since this construction uses only invariantly defined
geometric structures on $\HH$, the axial symmetry vector $\varphi^a$
is preserved, i.e.
\begin{equation}
  \mathcal{L}_X\varphi^a = 0\,.
\end{equation}
From the previous two equations and Eq.~(\ref{eq:zeta-defn}), it
follows that $\zeta$ is preserved as well: $\mathcal{L}_X\zeta = 0$.
Thus we construct the preferred coordinates $(\theta,\phi)$ as above
on each MOTS and then we simply take $X^a$ such that $\zeta$ (or
equivalently $\theta$) remains fixed.  We would still have the freedom
to add a shift in the $\varphi$ direction, but for non-spinning black
holes, we can choose the shift to be completely in the $\zeta$
direction.  In our case, it turns out that this construction leads to
a non-zero shift vector in the $\zeta$ direction.

\subsection{Fluxes, balance laws and multipole moments}
\label{subsec:fluxintro}

We conclude this section by summarizing the flux law for spacelike
dynamical horizons and the notion of multipole moments.  The reason
the area of a horizon increases is, of course, due to in-falling
radiation and matter. The same applies to angular momentum, mass and
higher multipole moments.  This can be seen as a ``physical process''
version of the first law of black hole thermodynamics. For spacelike
dynamical horizons it is possible to derive exact expressions for
these fluxes in full non-linear general relativity.  Since
$\HH$ is spacelike, it is equipped with a unit timelike normal
$\widehat{\tau}^a$, and each leaf of $\HH$ has a unit
spacelike normal $\widehat{r}^a$ tangent to $\HH$.  Then, a choice of null normals defined by $\HH$ is 
\begin{equation}
  \label{eq:normals-dh}
  \widehat{\ell}^a = \frac{1}{\sqrt{2}}\left(\widehat{\tau}^a + \widehat{r}^a\right)\,,\quad   \widehat{n}^a = \frac{1}{\sqrt{2}}\left(\widehat{\tau}^a - \widehat{r}^a\right)\,.
\end{equation}
This is analogous to Eq.~(\ref{eq:normals}), but the two choices are
different and related by a scaling.  Let $A_i$ and $A_f$ be the
initial and final areas respectively of a (not necessarily
infinitesimal) portion $\Delta\HH$ of a spacelike dynamical
horizon and let $\Delta R = R_f - R_i$ be the change in the area
radius.  Then, in vacuum spacetimes,
\begin{equation}
  \label{eq:dhflux}
  \Delta R = \frac{1}{4\pi}\int_{\Delta \HH} \left( \widehat{\sigma}_{ab}\widehat{\sigma}^{ab} + 2\widehat{\xi}_a\widehat{\xi}^a \right) N_R\,d^3V\,.
\end{equation}
Here $\widehat{\sigma}$ is the shear of the outgoing null normal
$\widehat{\ell}^a$,
$\widehat{\xi}^a = \twoq^{ab}\widehat{r}^c\nabla_c\widehat{\ell}_b$,
and $N_R$ is a suitable lapse function.  The integrand in this
expression is manifestly positive definite.  The important point here
is that we have identified the shear and the vector $\widehat{\xi}^a$
as the relevant fields which carry energy across $\HH$.  We have
already written the shear as a complex field $\sigma$ of spin weight
2, and we can similarly write $\widehat{\xi}^a$ as a complex field
$\widehat{\xi} = \widehat{\xi}^am_a$ of spin weight 1.  The
identification of $\sigma$ as an important part of the energy flux is
similar to the flux across null surfaces \cite{Hawking:1972hy}; see
also
\cite{Booth:2006bn,Booth:2003ji,OSullivan:2014ywd,OSullivan:2015lni,Prasad:2020xgr}.
The presence of the additional spin weight 1 field $\widehat{\xi}$
occurs because we are here dealing with non-null surfaces.  It is also
worth noting that $\widehat{\xi}$ becomes numerically difficult to
calculate as $\HH$ approaches equilibrium and becomes null
($\widehat{\tau}^a$ and $\widehat{r}^a$ are ill-behaved in the limit).
Below we shall study the decomposition of $\sigma$ into modes of spin
weight 2, and their time evolution. Note that we shall use
$\ell^a$ defined in Eq.~(\ref{eq:normals}) and not
Eq.~(\ref{eq:normals-dh})
for computing the shear and $\xi$.

Also of importance for us in this paper will be the notion of
multipole moments \cite{Ashtekar:2004gp} for axisymmetric horizons,
analogous to the well known Geroch-Hansen multipole moments at
infinity \cite{Geroch:1970cd,Hansen:1974zz}.  These were first defined
for isolated horizons where it can be shown that the two-dimensional
scalar curvature $\mathcal{R}$ and the rotational 1-form $\omega_a$
characterize the geometry of an isolated horizon.  Thus, by
considering multipole moments of these fields, one can characterize
the horizon geometry completely with a set of multipole moments (see
also \cite{Owen:2009sb} for an alternate set of moments).  These
multipole moments continue to be useful even in dynamical cases
\cite{Ashtekar:2013qta}.  Specifically, since we are dealing with
non-spinning black holes, we only need to consider $\mathcal{R}$,
which lead to the mass multipole moments of an axisymmetric MOTS
$\Surf$:
\begin{equation}
  I_l = \frac{1}{4}\oint_\Surf\mathcal{R}Y_{l,0}(\zeta)\,d^2V\,.
\end{equation}
Here $\zeta$ is the invariant coordinate defined in
Eq.~(\ref{eq:zeta-defn}), and $Y_{l,0}$ is the corresponding spherical
harmonic.  It is clear that the lowest moment $I_0$ is just a
topological invariant, and for spherical topology
$I_0=\sqrt{\pi}$. Furthermore, $I_1$ can be shown to vanish
identically from the definition of the coordinate system (in effect
these invariant coordinates automatically place us in the center of
mass of the system).  Non-trivial information is obtained from $l=2$
onwards, i.e. from the mass quadrupole, octupole etc.

\section{The spectrum of the stability operator}
\label{sec:stability}

In this section we describe the spectrum of the stability operator for
the various horizons.  We consider mostly $L_\Sigma$, and $L^{(-n)}$
briefly (both have qualitatively similar features).  We will break up
the discussion into three parts considering in turn the principal
eigenvalue, a selection of the next eigenvalues, and then finally a
statistical analysis of the higher eigenvalues.

\subsection{The principal eigenvalues}

\begin{figure*}
  \includegraphics[width=0.45\linewidth]{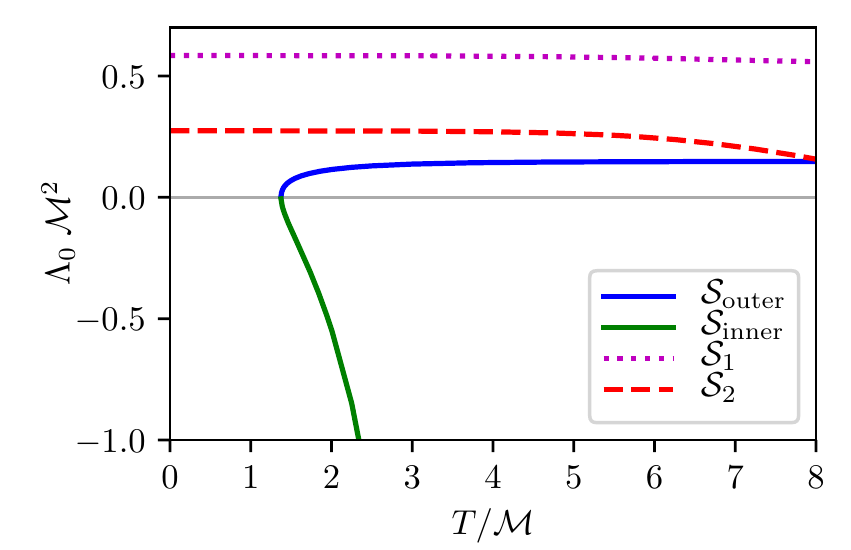}
  \includegraphics[width=0.45\linewidth]{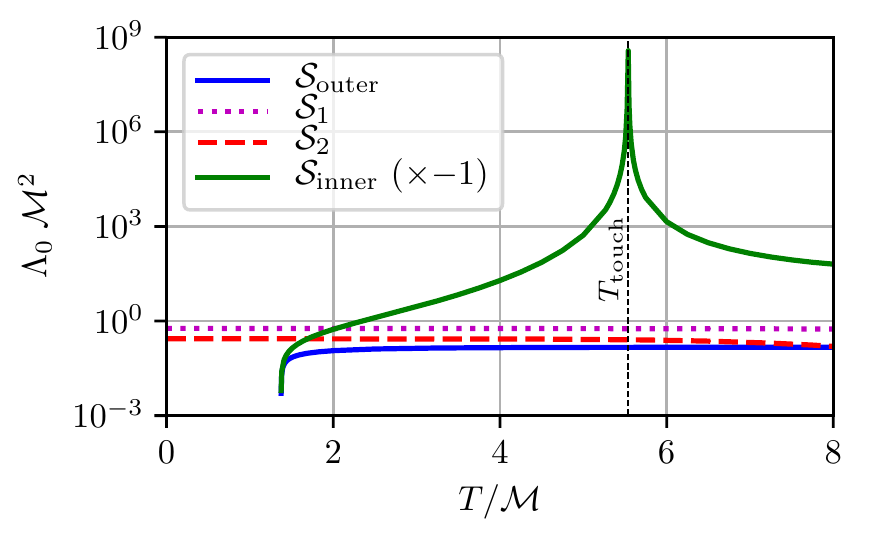}
  \caption{The principal eigenvalue $\Lambda_0$ of $L_\Sigma$ for all
    the horizons.  Except $\Sin$, all the horizons have positive
    $\Lambda_0$.  This is easily seen in the left panel.  For $\Sin$,
    $\Lambda_0$ shows a cusp at $\ttouch$. This is shown in the
    right panel on a logarithmic scale (we plot $-\Lambda_0$ for
    $\Sin$ because of the logarithmic scale).}
  \label{fig:stability-principal}
\end{figure*}

\begin{figure}
  \centering
  \includegraphics[width=0.9\columnwidth]{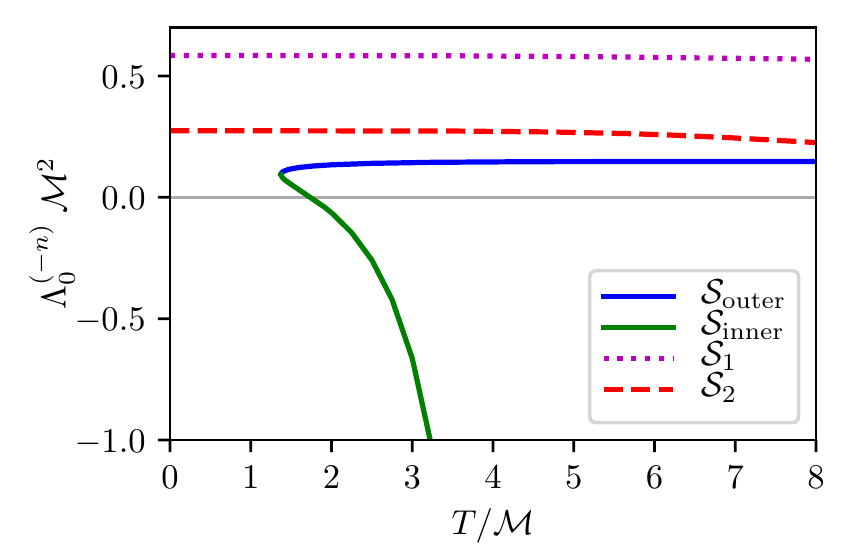}
  \caption{The principal eigenvalue for $L^{(-n)}$ for the various
    horizons. These values turn out to be somewhat larger than the
    corresponding values for $L_{\Sigma}$. Thus, the bifurcation
    between $\Sin$ and $\Sout$ occurs at a positive value of
    $\Lambda_0$.  Thus, $\Sin$ has positive principal eigenvalue for a
    short duration, and it does not cease to exist when $\Lambda_0$
    crosses zero. }
  \label{fig:stability-principal-null}
\end{figure}

Beginning with the principal eigenvalues of $L_\Sigma$, we have
already mentioned that for $\Sone$, $\Stwo$, and $\Sout$, $\Lambda_0$
is always positive.  $\Sout$ is born with $\Lambda_0=0$, but it
immediately becomes positive and remains so.  At early times for
$\Sone$ and $\Stwo$, and at late times for $\Sout$, two things happen:
i) the flux $|\sigma|^2$ is small and thus the differences between
$L_\Sigma$ and $L^{(-n)}$ are small. ii) The scalar curvature
$\mathcal{R}$ has only small variations, and thus the spectrum of
$L^{(-n)}$ is almost the same as that of the Laplacian on a round
sphere, with a shift corresponding to the value of the curvature.
Thus, in this limit where $\mathcal{R} \approx 2/R^2 = 1/2\Mirr^2$
with $R$ being the area radius, and $\Mirr$ being the irreducible
mass, the eigenvalues are labeled by two quantum numbers $(l,m)$ and
will be approximately\footnote{Note that, for $l=0$, this expression
  can be justified without assuming spherical symmetry
  (cf. Appendix~\ref{sec:linear_lamnbda_o}).}
\begin{equation}
  \label{eq:spectrum-round}
  \Lambda_{l,m} \approx \frac{1}{4\Mirr^2}\left(1+l(l+1)\right)\,.
\end{equation}
The state $l$ is $(2l+1)$-fold degenerate in this limit.
In the general but still axisymmetric case, the degeneracy between
states of different $|m|$ is broken, with $m$ being the label for the
angular modes.  The fundamental angular mode $m=0$ will in general not
be degenerate, while we find a $2$-fold degeneracy ($\pm m$) for the
higher modes with $m \neq 0$ due to axisymmetry.  The principal
eigenvalues of $L_\Sigma$ are shown in
Fig.~\ref{fig:stability-principal} for all the horizons, whereas
Fig.~\ref{fig:stability-principal-null} shows the principal
eigenvalues of $L^{(-n)}$.  The main difference with $L_\Sigma$ is
that $\Sin$ has positive principal eigenvalue for a short duration.

Most of this analysis does not apply to $\Sin$.  Just like $\Sout$,
the inner horizon $\Sin$ is born with $\Lambda_0=0$.  However unlike
$\Sout$, it becomes \emph{negative} thereafter.  $\Sin$ is therefore
unstable -- it is not strictly stably outermost and there are thus no
outward deformations which could make it strictly untrapped.  It is
also far too distorted for Eq.~(\ref{eq:spectrum-round}) to be even a
rough approximation to its spectrum.  We see from the right panel of
Fig.~\ref{fig:stability-principal} that $\Lambda_0$ for the inner
horizon apparently diverges to $-\infty$ at $\ttouch$ where it has a
cusp (though of course we cannot really prove this numerically).
\begin{figure}
  \centering
  \includegraphics[width=0.9\columnwidth]{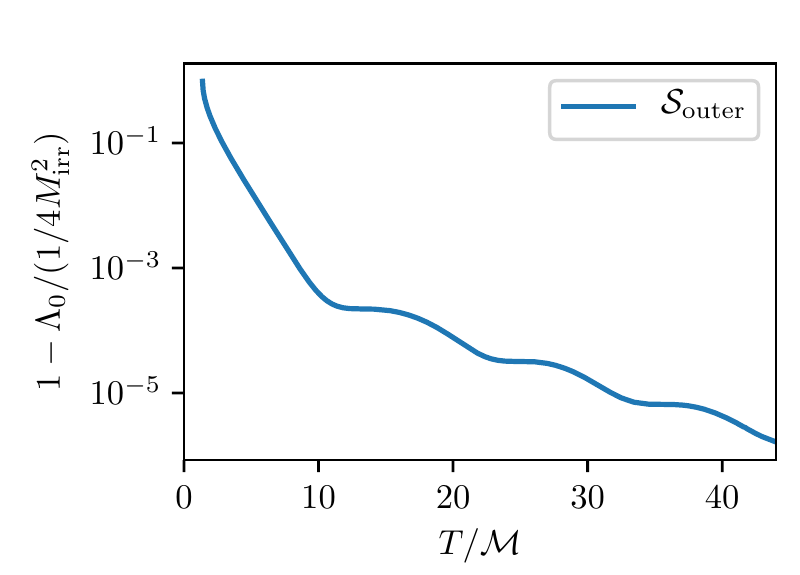}
  \caption{%
    Comparison of $\Lambda_0$ with the perturbative result
    \eqref{eq:perturbative_principal}.  Around $T\sim10\,\MM$, the
    curve changes from a steep to a more shallow exponential decay.  }
  \label{fig:stability-perturbative}
\end{figure}

This divergence, if it indeed exists, can be understood as follows.
Given the structure of the stability operator, it is tempting to
interpret it as the Hamiltonian of a quantum particle living on a
sphere.  The Laplacian is the analog of the kinetic energy while the
other terms in $L^{(-n)}$ and $L_\Sigma$ can be viewed as a potential.
The ground state energy is then the analog of $\Lambda_0$.  This
analogy can be extended also for spinning black holes where $\omega_a$
is non-vanishing \cite{Jaramillo:2014oha,Jaramillo:2015twa}.  Then,
the ground state energy will diverge to $-\infty$ only if the
potential also diverges to $-\infty$.  Of course, just because the
potential diverges at a point does not mean the ground state energy
also diverges; the hydrogen atom being the classic example.  Whether
or not $\Lambda_0\rightarrow -\infty$ depends on the details of how
$\mathcal{R}$ diverges at the cusp\footnote{This can be studied using
  the Lieb-Thirring inequality which relates the negative eigenvalues
  to the negative part of the potential (see
  e.g. \cite{lieb-thirring}).  In quantum mechanics, this inequality
  plays a critical role in mathematically proving that matter is
  stable.}.  For $L^{(-n)}$, the potential is just $\mathcal{R}/2$,
which is partially negative for $\Sin$ near the cusp
\cite{Pook-Kolb:2018igu}, and it diverges at $\ttouch$.  For
$L_\Sigma$, the potential also contains $|\sigma|^2$ which complicates
matters somewhat.  However, since $|\sigma|^2$ is non-negative and
comes with a negative sign, we see that the potential will still
diverge.  A detailed investigation of this mathematical question will
take us too far afield from the goals of our numerical study here, and
thus we will postpone this to future work.

There is one result for $\Hout$ that will be important for us later,
namely its approach to equilibrium.  Having computed $\Lambda_0$ for
$\Hout$ at all times, we can ask how it approaches the equilibrium
result of Eq.~(\ref{eq:spectrum-round}).  For $l=0$, we must have
$\Lambda_0\rightarrow \frac{1}{4\Mirr^2}$ at late times whence we can
compare $4\Mirr^2\Lambda_0$ with unity.  This is shown in
Fig.~\ref{fig:stability-perturbative} on a logarithmic scale.  We see
clearly a steep initial decay just after $\tbifurcate$, followed by a
shallower decay and oscillations.  We observe a transition between the
two regimes at $\approx 10\MM$.  This is our first encounter with this
kind of behavior, and we shall see this same pattern repeatedly
numerous times in this paper.  We shall study this behavior
quantitatively in detail for other geometric fields on $\Hout$ in the
following sections.

\subsection{Low eigenvalues}

\begin{figure*}
  \centering
  \includegraphics[width=0.45\linewidth]{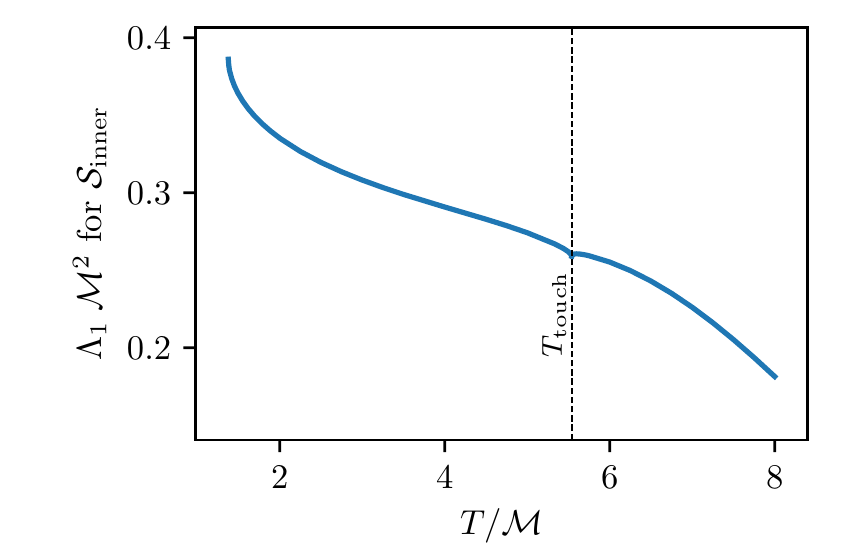}
  \includegraphics[width=0.45\linewidth]{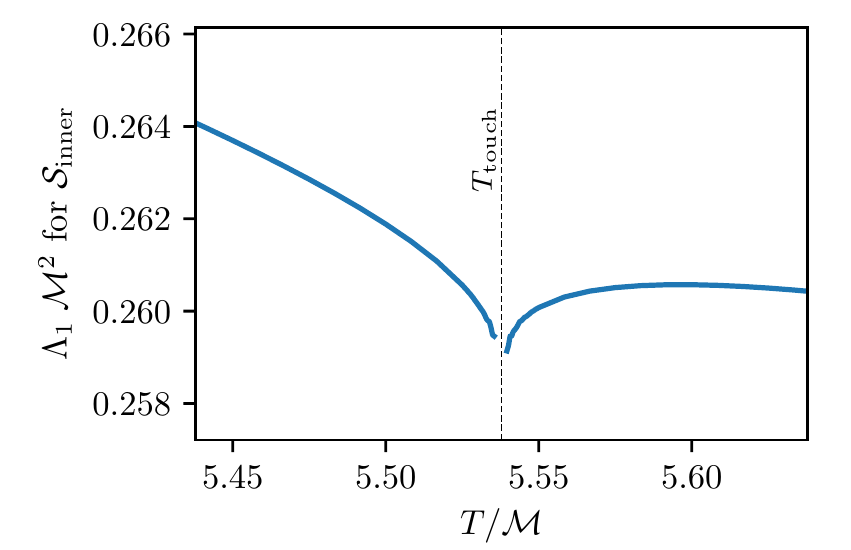}
  \caption{The second eigenvalue $\Lambda_1$ for $\Sin$ with angular
    mode $m=0$. The second
    panel shows a close-up near $\ttouch$.  The graph appears to show
    cusp-like behavior at $\ttouch$.  }
  \label{fig:stability-second}
\end{figure*}

\begin{figure}
  \centering
  \includegraphics[width=0.95\columnwidth]{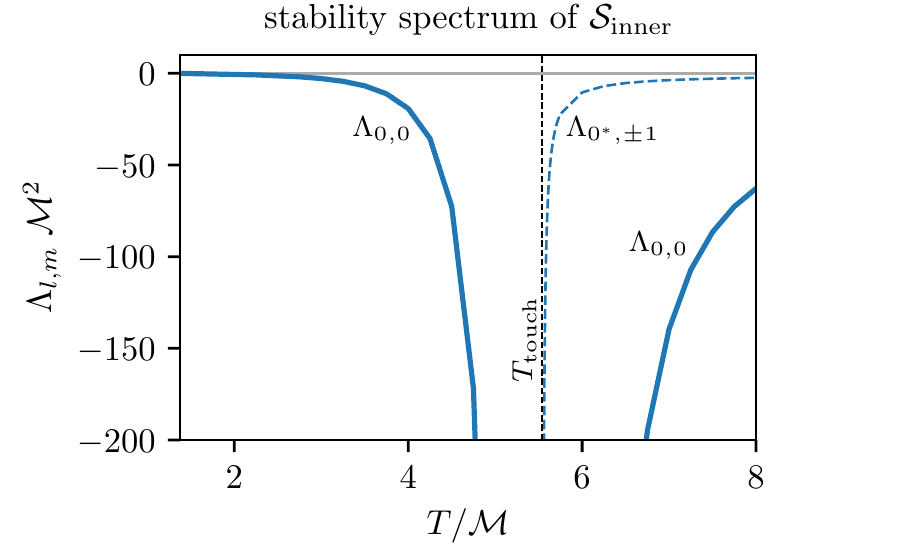}
  \caption{%
    The negative eigenvalues for $\Sin$.
    After $\ttouch$, two new (degenerate) negative eigenvalues
    appear for the $m=\pm1$ angular modes.
  }
  \label{fig:stability-negative}
\end{figure}

For $\Sone$, $\Stwo$ and $\Sout$, all the higher eigenvalues must be
positive since $\Lambda_0>0$.  Also for $\Sin$, apart from
$\Lambda_0$, all other eigenvalues must be positive till $\ttouch$.
The reason is that at $\tbifurcate$, $\Lambda_0=0$ and all the other
eigenvalues are positive definite.  Since the evolution is smooth, the
other eigenvalues must remain positive as long as $\Sin$ exists.  If
any of these eigenvalues were to cross zero, $\Sin$ would cease to
exist.  Fig.~\ref{fig:stability-second} therefore shows the next
eigenvalue $\Lambda_1$.  It turns out to be positive with possibly a
cusp at $\ttouch$.  This is shown in the second panel of
Fig.~\ref{fig:stability-second}.  We see that the graph of $\Lambda_1$
as a function of time appears to be forming a cusp at $\ttouch$,
though we are not numerically able to resolve this.  The precise value
of $\Lambda_1$ at the cusp is of interest.  If this were to be
negative, then it means that $\Lambda_1$ vanishes before $\ttouch$ and
therefore $\Sin$ does not exist near the cusp.  This seems unlikely
since we find $\Sin$ very shortly after $\ttouch$.  It seems more
reasonable to assume that $\Sin$ exists at all times around $\ttouch$
and our numerical methods are not able to locate it.  This implies
that the value of $\Lambda_1$ at $\ttouch$ should be non-negative.  It
would be interesting to prove (or disprove) this conjecture. In any
event, $\Lambda_1$ is still far from vanishing at the last time before
$\ttouch$ when it is located, indicating that it must exist for at
least a short time longer.  Similarly, at the first time it is located
after $\ttouch$, $\Lambda_1$ is similarly positive indicating that it
must have existed for at least a short time earlier.
Interestingly, the two lowest degenerate eigenvalues with angular
modes $m=\pm 1$ are positive before $\ttouch$, while after $\ttouch$
the lowest $m=\pm 1$ eigenvalues become negative.  We chose to label
these as new eigenvalues without relabeling the higher $m=\pm1$
ones. That is, instead of the usual
$\Lambda_{1,1} < 0 < \Lambda_{2,1} < \ldots$ we assign the labels
$\Lambda_{0^*,1} < 0 < \Lambda_{1,1} < \ldots$.  This is shown in
Fig.~\ref{fig:stability-negative}.  These $\Lambda_{0^*,\pm1}$
eigenvalues are seen to increase much more rapidly than $\Lambda_0$
itself but, as far as we are able to track $\Sin$, none of these
eigenvalues cross zero and $\Sin$ continues to exist.

\subsection{Global behavior of the spectrum}

\begin{figure*}
  \centering
  \includegraphics[width=0.45\linewidth]{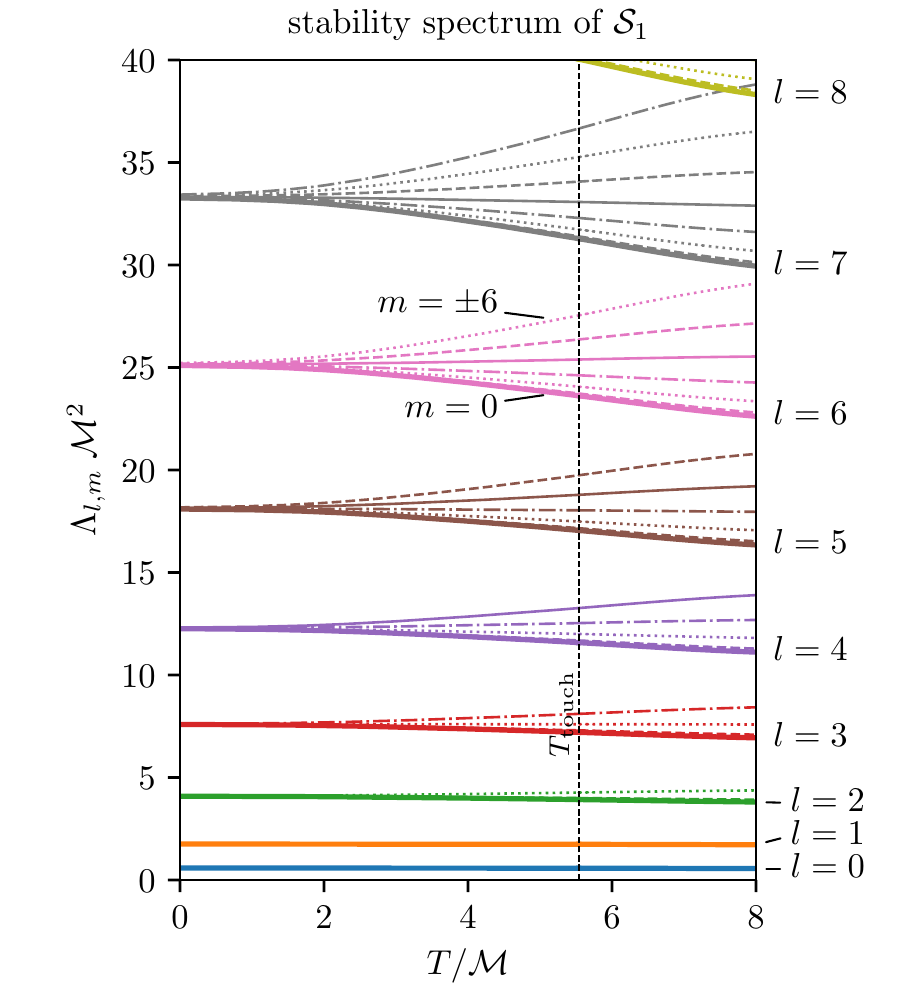}
  \includegraphics[width=0.45\linewidth]{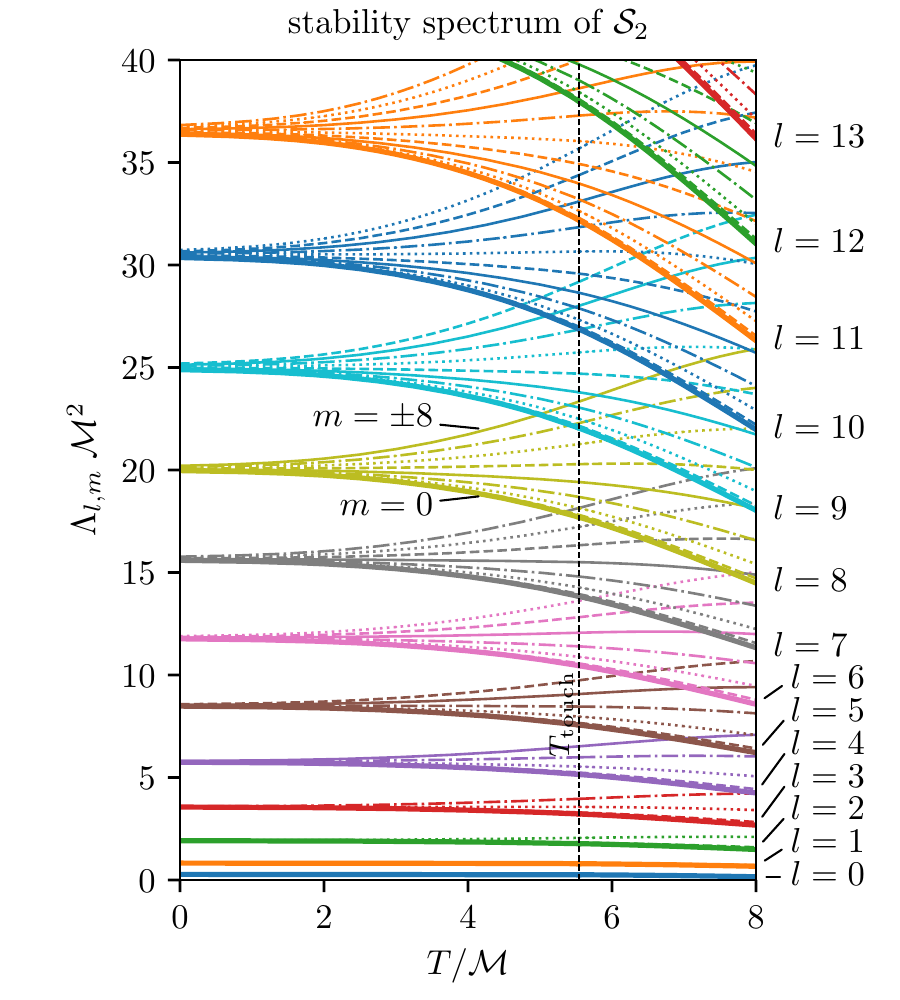}\\
  \includegraphics[width=0.45\linewidth]{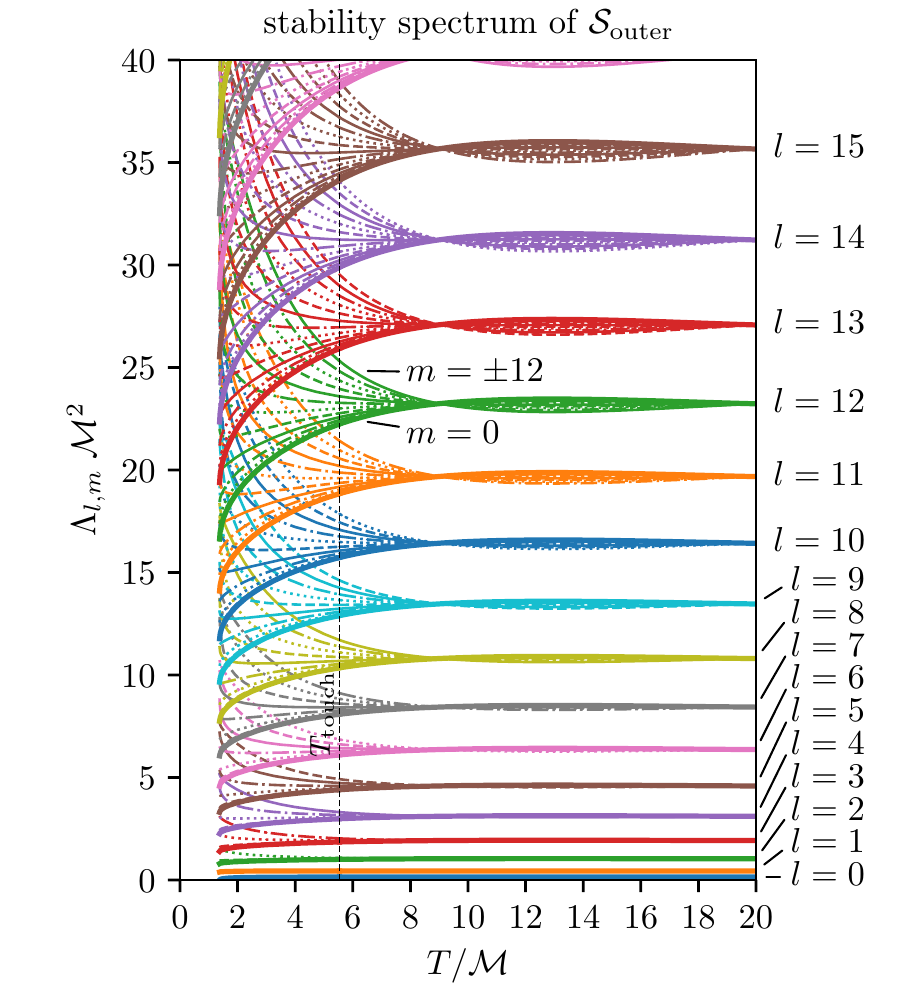}
  \includegraphics[width=0.45\linewidth]{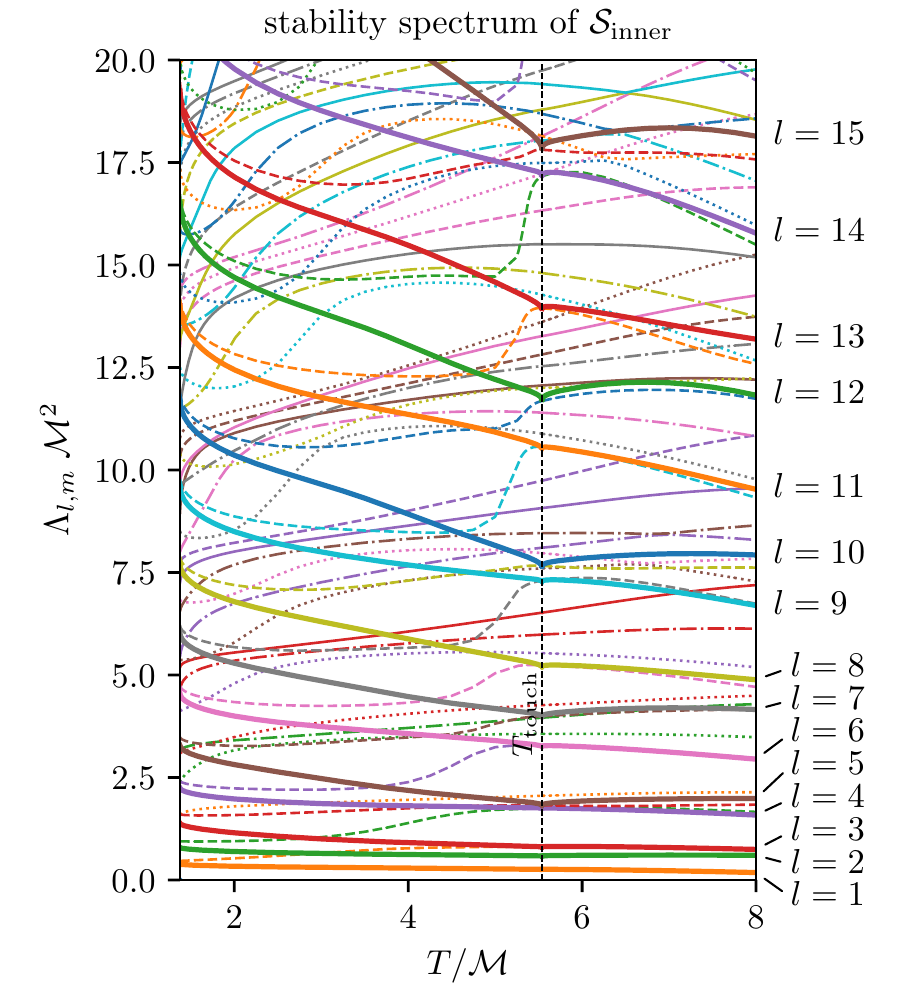}
  \caption{The stability spectrum of the various horizons.  As
    expected, $\Sone$ and $\Stwo$ (top two panels) start off with a
    simple spectrum corresponding to Eq.~(\ref{eq:spectrum-round}) and
    become more complicated near the merger.  The spectrum for $\Sout$
    in the bottom-left panel shows the opposite behavior.  The bottom
    right panel shows the positive part of the spectrum for $\Sin$.}
  \label{fig:stability-spectrum}
\end{figure*}

The higher eigenvalues of $L_\Sigma$ are shown\footnote{The spectrum
  of $L^{(-n)}$ has similar global properties, except that we obtain
  slightly larger values corresponding to $|\sigma|^2$, and in
  accordance with the general results in \cite{Andersson:2007fh}.  We
  have chosen to show just the principal eigenvalue, cf. Fig.~
  \ref{fig:stability-principal-null}.} in
Fig.~\ref{fig:stability-spectrum}. The top panels show the spectra for
$\Sone$ and $\Stwo$. At early times we have the behavior predicted by
Eq.~(\ref{eq:spectrum-round}).  The larger black hole, i.e. $\Stwo$,
has smaller eigenvalues for the same value of $l$. A multiplet
structure is apparent here.  As we get closer to the merger, the
states with different $m$ are no longer degenerate, analogous to the
splitting of energy levels of a quantum system in an external field.
The states with $\pm m$ remain degenerate due to axisymmetry.  For
generic configurations (including spins, non-zero orbital angular
momentum etc.), this symmetry would then not be present and the
$\pm m$ states would not be degenerate.

As we approach $\ttouch$, the energy levels are seen to cross and it
becomes more difficult to distinguish the states with different $l$,
though the multiplet structure with splitting can still be identified.
The apparent horizon has the opposite behavior.  It approaches this
simple spectrum at late times when it settles down to a Schwarzschild
black hole. The multiplet structure here is again apparent.

\begin{figure}
  \centering
  \includegraphics[width=0.95\columnwidth]{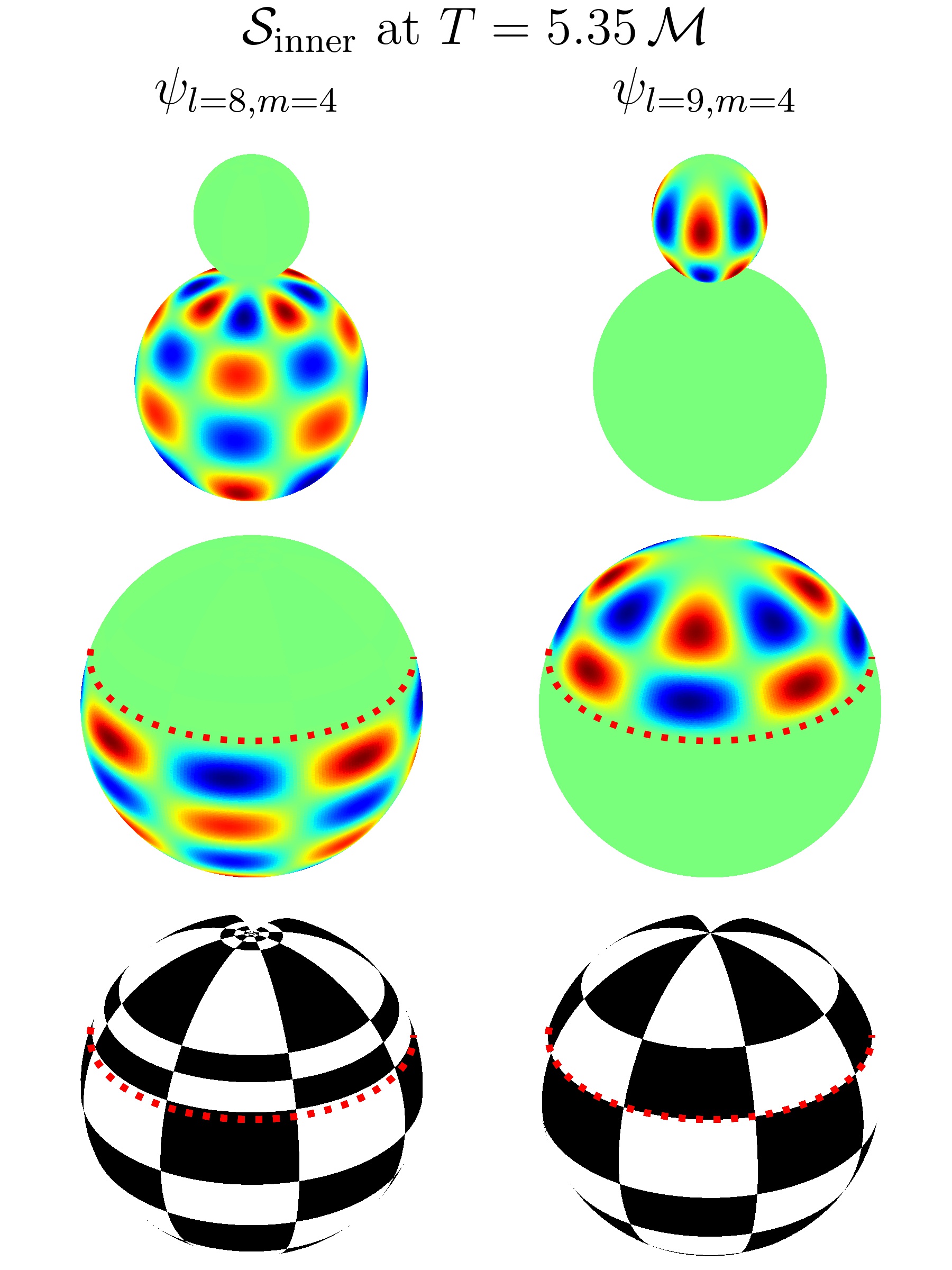}
  \caption{%
    Visualization of two eigenfunctions $\psi_{l,m}$ of $L_\Sigma$ for $\Sin$
    at a time $T=5.35\,\MM$ before $\ttouch\approx5.53781\,\MM$.
    The top row shows the function values in blue ($<0$) and red
    ($>0$) on $\Sin$, while the lower two rows show the function
    values and sign changes, respectively, as functions of
    $(\theta,\phi)$ on a sphere. The dotted line indicates the
    $\theta$-coordinate of the ``waist'' visible in the first row.
    From the bottom row it is clear that the extended green regions
    are only close to zero but still contain structure.
  }
  \label{fig:nodal-lines}
\end{figure}

The inner horizon $\Sin$ apparently shows no such simplicity.
Nevertheless, some spectroscopy-like analysis seems possible. In
particular, a transfer of states between different multiplets seems to
happen, with a migration of states from $l \rightarrow l+2$. This can
be understood in terms of tidal coupling. Specifically, at around
$\tname\sim 3\,\MM$, $\Sin$ is sufficiently deformed.  It structures
itself into two well identified lobes that ultimately pinch at
$\ttouch$. The system starts to effectively behave as a binary,
dramatically illustrated by the eigenfunctions which situate
themselves in either one or the other lobe (illustrated in
Fig.~\ref{fig:nodal-lines}). 
The two components of this ``quasi-binary'' interact tidally ($l=2$)
inducing this coupling in the spectrum levels.

In summary, this kind of non-trivial coupling between levels results
in a completely different multiplet restructuring after $\ttouch$
(e.g. the two lowest multiplets are singlets, as a consequence of the
loss of states to higher levels).  Globally, there turns out to be a
further complexity for the inner horizon that suggests the need to
resort to other systematic tools to probe its underlying structure.
Looking further ahead to future work when we consider more generic
configurations without axisymmetry, the spectrum will be complex and
yet more complicated.  It will not be possible to investigate each
eigenvalue in detail.  We must then resort to a statistical analysis
of the spectrum, from which we can extract valuable information.  The
remainder of this section can be seen as a precursor to the more
complicated case.

\subsubsection{Crossing of energy levels}

Still in a spectroscopic spirit, a clearly evident feature of the
spectra shown in Fig.~\ref{fig:stability-spectrum}, including that of
$\Sin$, is the crossing of eigenvalue levels.  This is very
significant, since it is not the generic situation for real
self-adjoint operators (of the class we are studying) depending on a
single parameter, time $t$ in our case. The variation of the
Hamiltonian with time typically leads to level repulsion, whereas
level-crossing requires two parameters \cite{Berry81}. This can be
accounted for in terms of the
corresponding classical dynamics, if the operator is understood as a
classical Hamiltonian on a phase space.  It turns out that for generic
classical Hamiltonian systems, namely non-integrable (or chaotic in
rough terms), level-crossing translates into an over-determined
condition which generically admits no solution if only one parameter
is available. As a result, eigenvalues repel, something that
quantum-mechanically corresponds to coupling of the levels and the
impossibility of defining quantum numbers.


On the contrary, when the underlying classical motion is integrable,
the eigenvalue curves indeed can
(quasi\nobreakdash-)cross\footnote{Actual crossing requires a stronger
  condition, namely separability, whereas in general integrable
  systems level lines can approach to extremely narrow separations but
  can then ultimately repel \cite{Berry81}.}.  Levels do not interact
and evolve independently, quantum numbers can be tracked and
clustering can happen due to the absence of level repulsion.  In our
present case, the corresponding classical system is not only
integrable, but our problem is actually separable\footnote{An
  interesting consequence of the separability of our eigenvalue
  problem, as a consequence of axisymmetry, is the crossing of nodal
  lines of the eigenfunctions. This is not the generic situation even
  for integrable system (c.f. e.g. \cite{Berger03}), and follows from
  separability in two-dimensions in an orthogonal coordinate
  system. This is illustrated in Fig.~\ref{fig:nodal-lines} for two
  eigenfunctions of $\Sin$.}  as a consequence of axisymmetry.  The
latter is a stronger (non-generic) feature that implies integrability
\cite{Berry81}.  From this perspective, nothing distinguishes $\Sin$
from the other horizons.  In summary, for the four spectra shown in
Fig.~\ref{fig:stability-spectrum}, level-crossing is a strong
indication of classical integrability and in our case a confirmation
of the a priori knowledge about the separability of the system.

\subsubsection{Spectrum statistics}

The spectrum of a given MOTS stability operator is of course purely
deterministic and can be efficiently calculated numerically.  The
underlying system, black holes in standard classical general
relativity, do not have any quantum aspects. However, we have found it
useful to think of the spectral problem as being associated with the
Hamiltonian of a quantum particle living on the MOTS.  We shall now
push this analogy further to the higher eigenvalues and borrow
techniques from quantum mechanics.  In the present self-adjoint
setting the operators $L^{(-n)}$ and $L_\Sigma$ can be seen (cf. sec
4.4. in \cite{Jaramillo:2014oha}) as the quantum Hamiltonian $\hat{H}$
corresponding to a classical Hamiltonian function
$H(p,q)=q^{ab}p_ap_b + \frac{1}{2}\frac{}{}\mathcal{R}(q)$ on the
cotangent bundle $T^*\mathcal{S}$.  Much insight can be gained then
into the actual MOTS spectrum from semi-classical considerations
connecting the quantum system defined by $\hat{H}$ to the underlying
classical Hamiltonian system
\cite{Berry81,bohigas1986,berry1987,Eckha88,WimbergerBook}. Tools and
concepts from the study of quantum chaos will be adapted to the present
MOTS setting.  Different eigenvalue-level statistics can be devised to
address distinct aspects of the spectrum.  We will focus here on the
small scale aspects of the spectrum, i.e. the interaction between
adjacent levels.

For the higher eigenvalues, a statistical perspective on the
distribution of eigenvalues can reveal important structural features
of the underlying geometric object. This approach parallels the
research program initiated by Wigner \cite{Porter65} to undertake the
understanding of the spectral properties of complex heavy nuclei in
terms of statistical ensembles, leading to Dyson's random-matrix
models \cite{dyson-1,dyson-2,dyson-3}.  Later, these tools have been
also systematically employed in the setting of quantum chaos,
exploring the subtle interplay between the quantum and the underlying
semi-classical system.  Here we will focus on the application to our
spectra of a short-range correlation in the spectrum, namely the
`nearest neighbor spacing distribution' $P(S)$ which we describe
shortly.  This spectral statistic accounts for the fine-scale
structure of the spectrum and in particular it is sensitive to the
clustering or repulsion between the energy levels.

An important point is a need to remove ``trivial'' degeneracies due to
symmetries. In our case these degeneracies correspond to the $\pm m$
degeneracy.  We do not want the distribution $P(S)$ to be dominated by
this degeneracy, and thus they must be removed at the very start of the
analysis.  Eigenvalues can then be ordered as
\begin{equation}
\label{eq:lambda_ordering}
\Lambda_o < \Lambda_1 \leq \Lambda_2 \leq \ldots \leq \Lambda_n \leq \ldots\,, 
\end{equation}
where the non-degeneracy of $\Lambda_o$ has been taken into account.

Prior to the introduction of spectral statistics, we perform a
normalization of the spectrum by setting its average level density to
unity.  Specifically, we first introduce a function $N(\Lambda)$
counting the number of eigenvalues $\Lambda_i$ below a certain value
$\Lambda$ as
\begin{equation}
  N(\Lambda) = \sum_i \Theta(\Lambda - \Lambda_i) \,,
\end{equation}
where $\Theta(y) = 1- H(y)$, with $H(y)$ the Heaviside function.  The
counting function $N(\Lambda)$ has a staircase structure.  The level
density (density of states) is then defined as
\begin{equation}
  \rho(\Lambda) = \frac{dN}{d\Lambda} \,.
\end{equation}
We can write $N(\Lambda)$ as
\begin{equation}\label{eq:N_counting}
  N(\Lambda) = N_\mathrm{av}(\Lambda) + N_\mathrm{fl}(\Lambda) \,. 
\end{equation}
Here $N_\mathrm{av}(\Lambda)$ is a monotonically increasing smooth
function; it is the secular part of $N(\Lambda)$ interpolating the
steps in $N(\Lambda)$.  $N_\mathrm{fl}(\Lambda)$ is the fluctuating
part accounting for the difference with respect to the secular
increase.  The ``unfolding'' of the spectrum is a ``rectification'' of
the latter such that secular level density is $1$. In particular, by
introducing $x=N_\mathrm{av}(\Lambda$), for the ``unfolded spectrum''
\begin{equation}
  x_i =  N_\mathrm{av}(\Lambda_i)\,,
\end{equation}
we obtain an average level density of unity in the new variable
\begin{equation}
  \rho_\mathrm{av}(x) = \frac{dN_\mathrm{av}}{dx} = \frac{dN_\mathrm{av}}{d\Lambda} \frac{d\Lambda}{dx}=
  \frac{dN_\mathrm{av}}{d\Lambda}\left(\frac{dN_\mathrm{av}}{d\Lambda}\right)^{-1}=1 \,.
\end{equation}
We focus here on the fine scale features in the spectrum, in terms of
the distribution of separations between adjacent eigenvalues in
Eq.~(\ref{eq:lambda_ordering}).  Nearest-neighbor spacings $S_i$ are
calculated in the unfolded spectrum as
\begin{equation}
  S_i = x_{i+1}- x_i\,.
\end{equation}
The probability of finding a spacing between $S$ and $S+dS$ is given
by $P(S)dS$ and, because of using the unfolded spectrum, the average
spacing $\langle S\rangle$ is unity:
\begin{equation}
  \label{eq:S_normalized}
  \langle S\rangle = \int P(S) S dS = 1\,.
\end{equation}
Since $P(S)$ measures the correlation between adjacent eigenvalues,
$P(S)$ is said to be a ``short-range level'' correlation measure.  We
shall calculate $P(S)$ for the stability spectrum and attempt to
interpret the result as a representative of a particular universality
class.  As a trivial example of such a universality class, consider
the so-called ``picket fence'' distribution (namely a Dirac delta)
centered at unity:
\begin{equation}\label{eq:picket_fence_1}
  P(S) =\delta(S-1)\,.
\end{equation}
It is clear that such a distribution characterizes a perfectly regular
spectrum.

More interestingly, for real Laplacian-like operators as in
Eqs.~(\ref{eq:Ln_selfadjoint}) and (\ref{eq:Ln_sigma}), $P(S)$
presents a universality behavior according to the type of classical
motion, `integrable' versus `chaotic', of the corresponding classical
Hamiltonian:
\begin{itemize}
\item[i)] {\em ``Integrable'' classical motion}: In this case we
  obtain a Poisson distribution
  \begin{equation}
    \label{eq:Poisson}
    P(S) =e^{-S}\,. 
  \end{equation}
  This corresponds to a distribution showing a tendency to cluster
  since $P(0)\neq 0$.  Moreover, the levels $\Lambda(t)$
  cross\footnote{They can actually repel at an exponentially small
    scale \cite{Berry81}.}. In particular, crossing happens for
  separable systems. The associated degeneracy is accounted by a
  non-vanishing $P(0)$ and quantum numbers can be assigned to levels
  in a straightforward manner.

\item[ii)] {\em ``Chaotic'' classical motion}: This is the so-called
  Wigner surmise~\footnote{\label{footnote:GOE-GUE-GSE}%
    Very
    interestingly, the Wigner surmise appears also in the setting of
    the Gaussian Orthogonal Ensemble (GOE) universality class in
    random matrices. More generally, the Bohigas-Giannoni-Schmit
    conjecture (cf. e.g. \cite{WimbergerBook}), the eigenvalues
    corresponding to a chaotic classical system obey the same
    universal statistics of level spacings as those Gaussian random
    matrices~\cite{dyson-1,dyson-2,dyson-3}. In particular, real
    time-reversal symmetric systems follow Gaussian Orthogonal
    Ensemble (GOE) statistics, whereas (complex) non-time-reversal
    symmetric Hamiltonians are associated with the Gaussian Unitary
    Ensemble (GUE). Other ``more exotic'' non-time-reversal systems,
    appearing for instance in spin systems, are related to the
    Gaussian Symplectic Ensemble (GSE). For completeness, we present
    here the universal $P(S)$ distributions for GUE and GSE statistics 
\begin{equation}
\label{eq:GUE-GSE}
P_{\mathrm{GUE}}(S) =\frac{32}{\pi}S^2 e^{-\frac{\pi 4 S^2}{\pi}} \,,\
P_{\mathrm{GSE}}(S) =\frac{2^{18}}{3^6\pi^3}S^3 e^{-\frac{64 S^2}{9\pi}}\, .
\end{equation}}:
  \begin{equation}
    \label{eq:Wigner}
    P(S) =\frac{\pi}{2}S e^{-\frac{\pi S^2}{4}}\,. 
  \end{equation}
  This behavior displays repulsion between eigenvalues since
  $P(0) = 0$.  The eigenvalue curves (generically) do not cross
  \cite{Berry81}, and therefore they do not degenerate.  Level
  crossing requires two parameters.  Therefore close levels couple and
  repel, with the strength of the coupling given by the minimum energy
  difference between the two repelling eigenvalue curves. No ``quantum
  numbers'' can be assigned to such levels.
\end{itemize}
It is important to keep in mind that any of this behavior becomes
evident only after the ``trivial'' degeneracies due to symmetries have
been eliminated.  Long-range correlations can be studied with other
spectral statistics (cf. e.f. \cite{WimbergerBook}), such as the
number variance $\Sigma(L)$ or the spectral rigidity $\Delta(L)$,
presenting also universality in certain regimes (small $L$ in this
case). We postpone this to a later study.

\begin{figure*}
  \begin{subfigure}{.33\linewidth}
    \includegraphics[width=\textwidth]{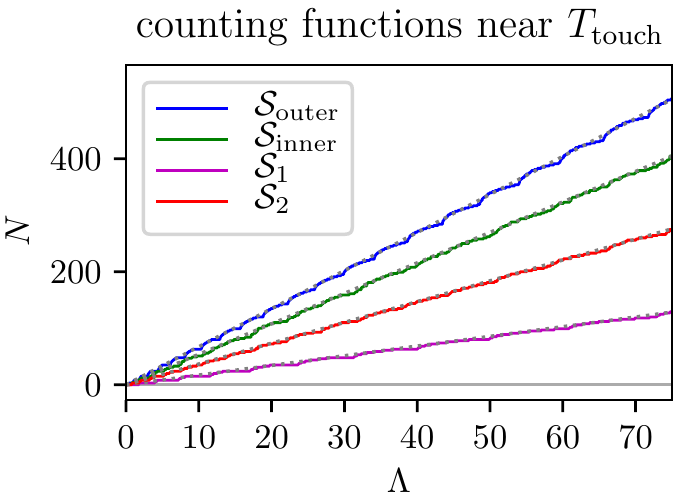}
    \caption{%
        Function $N$ for the four horizons.
        The dotted lines show Weyl's law.
    }
    \label{fig:sub-stability-N}
  \end{subfigure}%
  \hfill
  \begin{subfigure}{.33\linewidth}
    \includegraphics[width=\textwidth]{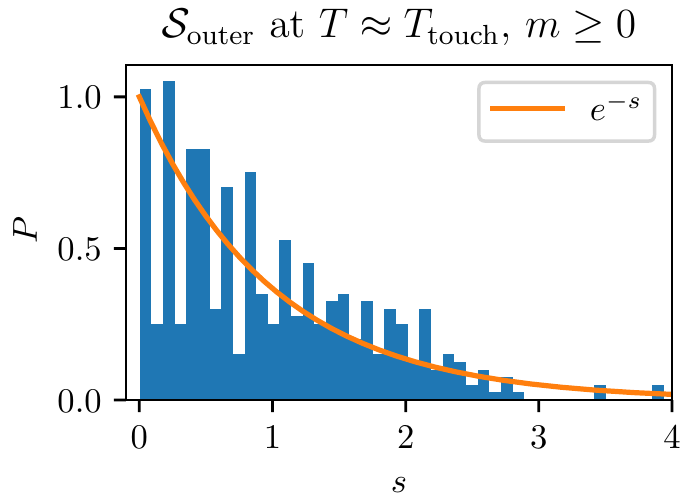}
    \caption{%
        Distribution $P$ for $\Sout$ showing a Poisson-like shape.
    }
    \label{fig:sub-stability-poisson-outer}
  \end{subfigure}%
  \hfill
  \begin{subfigure}{.33\linewidth}
    \includegraphics[width=\textwidth]{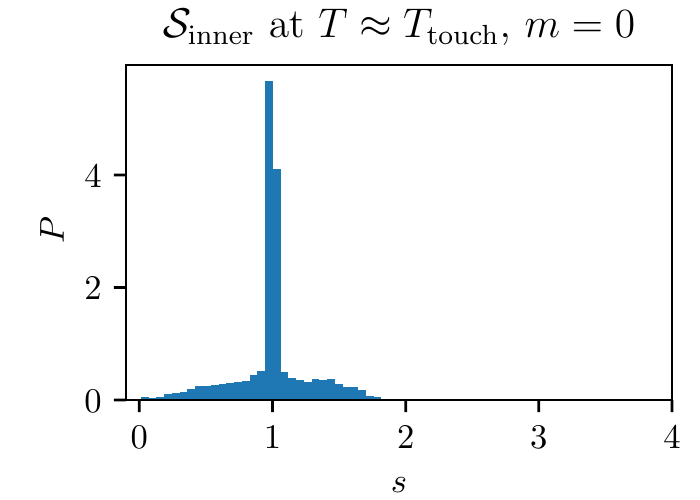}
    \caption{%
        ``Picket-fence'' distribution when considering a fixed $m$ spectrum.
    }
    \label{fig:sub-stability-picket}
  \end{subfigure}%
  \hfill
  \begin{subfigure}{.33\linewidth}
    \includegraphics[width=\textwidth]{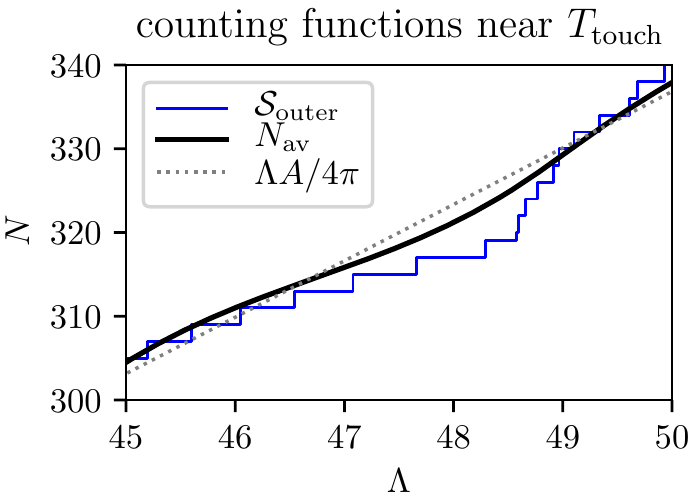}
    \caption{%
        Close-up of (a) showing $N_\text{av}$.
    }
    \label{fig:sub-stability-N-zoom}
  \end{subfigure}%
  \hfill
  \begin{subfigure}{.33\linewidth}
    \includegraphics[width=\textwidth]{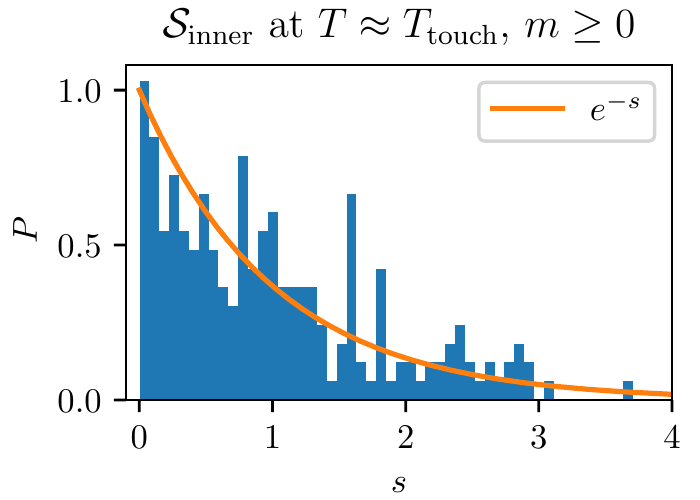}
    \caption{%
        Poisson-like distribution for $\Sin$.
    }
    \label{fig:sub-stability-poisson-inner}
  \end{subfigure}%
  \hfill
  \begin{subfigure}{.33\linewidth}
    \includegraphics[width=\textwidth]{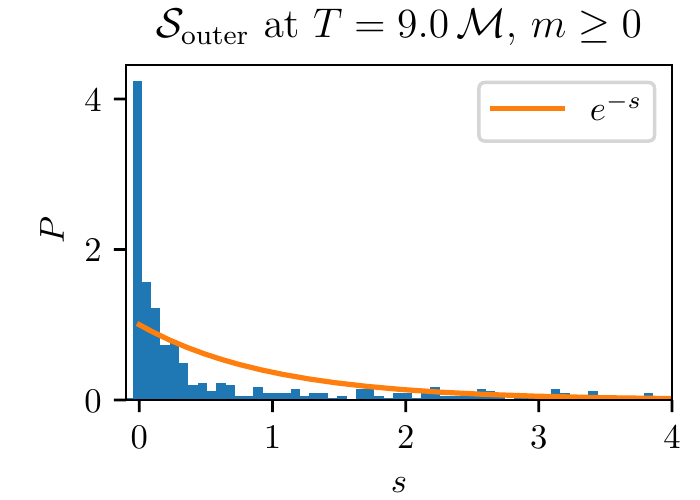}
    \caption{%
        ``Quasi-picket-fence'' at $T=9\,\MM$.
    }
    \label{fig:sub-stability-quasi-picket}
  \end{subfigure}
  \caption{%
    Construction and examples of the spectrum statistics.
    See text for details.
  }
  \label{fig:spectrum-stats}
\end{figure*}

We are now ready to apply the above formalism to the stability
spectrum.  We start by mapping the spectrum to the ``unfolded''
spectrum where the average spacing between neighboring levels is
normalized to $1$. For this we first determine the average
$N_{\mathrm{av}}(\Lambda)$ of the spectrum level-counting function
$N(\Lambda)$. Fig.~\ref{fig:spectrum-stats} shows in panel~(a) the
step-wise $N(\Lambda)$ 
for all four horizons at a time very close to $\ttouch$.  In
particular, we note the nice agreement with Weyl's law (see
Appendix~\ref{app:Weyl}) at large eigenvalues.

Then defining the unfolded levels as $x_i=N_{\mathrm{av}}(\Lambda_i)$,
we can construct the distribution of the nearest-neighbor distance
variable, $S_i = x_{i+1}- x_i$. First we notice that if only
eigenvalues with a fixed $m$ are considered, then we obtained a
perfectly regular distribution corresponding to a ``picket fence''
centered at $S=1$ given in Eq.~(\ref{eq:picket_fence_1}).  This case is
shown in panel~(c) of Fig.~\ref{fig:spectrum-stats}.  This is
non-generic behavior, resulting from axisymmetry where $m$ is the only
preserved quantum number for all times.  The distribution is dominated
by this degeneracy and we are not able to infer any relevant
non-trivial structure.  To fix this, consider now all eigenvalues with
the $\pm m$ symmetry removed. The resulting histogram for $P(S)$ is
shown in Fig.~\ref{fig:spectrum-stats}, panels~(b) and (e).  As
expected, a Poisson distribution is obtained for both $\Sout$ and
$\Sin$ despite their very different appearance in
Fig.\ref{fig:stability-spectrum}. This is a consequence of the
underlying classical integrability. The effect of level-crossing is
apparent in the non-vanishing value of $P(0)$, indicating the generic
occurrence of degeneracies.

Finally, we comment on the oscillations of the eigenvalues visible in
Fig.~\ref{fig:stability-spectrum}.  For example, near $T\approx 9\MM$,
we see from the bottom-left panel of the figure that the eigenvalues
with the same $l$ (but different $m$) are apparently almost
degenerate.  Remarkably at this time, the spectrum is in fact very
close to that of a round sphere -- the various oscillation modes of
the MOTS conspire near this time to produce a nearly round sphere for
a short duration.  Panel~(f) of Fig.~\ref{fig:spectrum-stats} shows
the distribution $P(S)$ at this time.  This is very close to a
quasi-picket-fence distribution centered at $S=0$ in.  As we shall
explain later, this behavior is consistent with the observed evolution
of the horizon multipoles in Fig.~\ref{fig:multipoles-common}.

Regarding $\Sin$, we note that the $P(S)$ statistic does not capture
many specific features of the spectrum. This includes, for example,
the multiplet reorganization between different levels, which is not a
short-correlation effect. Addressing this requires the implementation
of statistics for long-range correlations among spectrum levels, such
as the number variance $\Sigma(L)$ or the spectral rigidity
$\Delta(L)$, and will be done somewhere else.  Finally, the present
spectrum statistics analysis could have been anticipated from the a
priori knowledge of the system separability. The interest therefore
lies in providing a benchmark for future comparison with generic
binary mergers where separability will be lost and, presumably,
classical integrability will also disappear.

\section{Horizon shear and fluxes}
\label{sec:flux}

Paper I has provided a detailed understanding of how the area
increases.  Now we turn our attention to \emph{why} the area
increases, i.e. because of the in-falling flux of radiation (and
potentially matter fluxes if we had matter fields).  Recall here the
expression for the area flux given in Eq.~(\ref{eq:dhflux}).  There
are two contributions, the first being the familiar shear term.  This
is analogous to the well known outgoing radiation at least in the
sense that the shear is a field of spin weight 2.  It has been
observed to be closely correlated with the News tensor at null
infinity \cite{Prasad:2020xgr}.  The second term involving $\xi$ has
no corresponding counterpart at null infinity (this is not surprising
given that the dynamical horizon is not null). Being a vector field,
$\xi = \xi_am^a$ has spin weight $+1$.

\begin{figure}
  \centering    
  \includegraphics[width=0.9\linewidth]{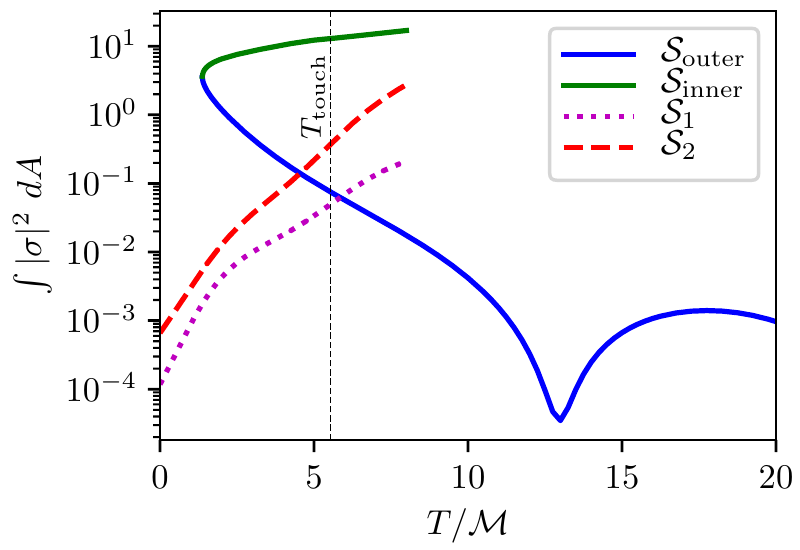}
  \caption{The integral of $|\sigma|^2 := \sigma_{ab}\sigma^{ab}$ for
    the outgoing normal $\ell^a$ given in Eq.~(\ref{eq:normals}). The
    dashed and dotted lines are for the individual horizons while the
    solid lines are for the two common horizons.  }
  \label{fig:shear-integral}
\end{figure}

The dominant term in the flux is the shear.  Let us therefore consider
the 2-dimensional integral of $|\sigma|^2$ over the various MOTSs; let
us call this the shear flux.  The result is shown in
Fig.~\ref{fig:shear-integral}.  The shear-flux increases for $\Sone$
and $\Stwo$, while it decreases for $\Sout$.  The dip in the
shear-flux for $\Sout$ near $T\approx 13\,\MM$ is because of an
oscillation in the dominant $l=2$ mode of the shear as we shall see
below.  This is to be compared with Fig.~10 of paper I showing the
corresponding dip in the plot of the rate of change of the area as a
function of time.  For the inner-common horizon $\Sin$, the shear-flux
increases rapidly in the beginning and soon reaches a plateau.  It is
noteworthy that there is no discontinuity across the merger when
$\Sin$ develops a cusp and then self-intersections.

Being a symmetric tracefree tensor, we expand $\sigma$ in spherical
harmonics of spin weight $+2$. We have already constructed in
Sec.~\ref{subsec:coordinates} a preferred coordinate system
$(\theta,\phi)$ which exploits the axisymmetry of the problem. These
coordinates can obviously also be used for our needs in this section,
i.e. expanding spin weight 2 fields.  For the complex scalar $\sigma$
we get
\begin{equation}
  \label{eq:shear-decomposition}
  \sigma(\theta,\phi,t) = \sum_{l=2}^{\infty}\sum_{m=-l}^{l}\sigma_{l m}(t){}_{2}Y_{l m}(\theta,\phi)\,.
\end{equation}
Here ${}_{2}Y_{l m}$ are spin-weighted spherical harmonics and
$\sigma_{l m}$ are the mode amplitudes.  This decomposition can be
carried out for all of the horizons in our problem, namely the two
individual and the two common horizons. Furthermore, since we have
explicit axisymmetry with $\sigma$ independent of $\phi$, we will only
have the $m=0$ modes and we will drop the index $m$ in
$\sigma_{l,m}$.

\begin{figure*}
  \centering    
  \includegraphics[width=\columnwidth]{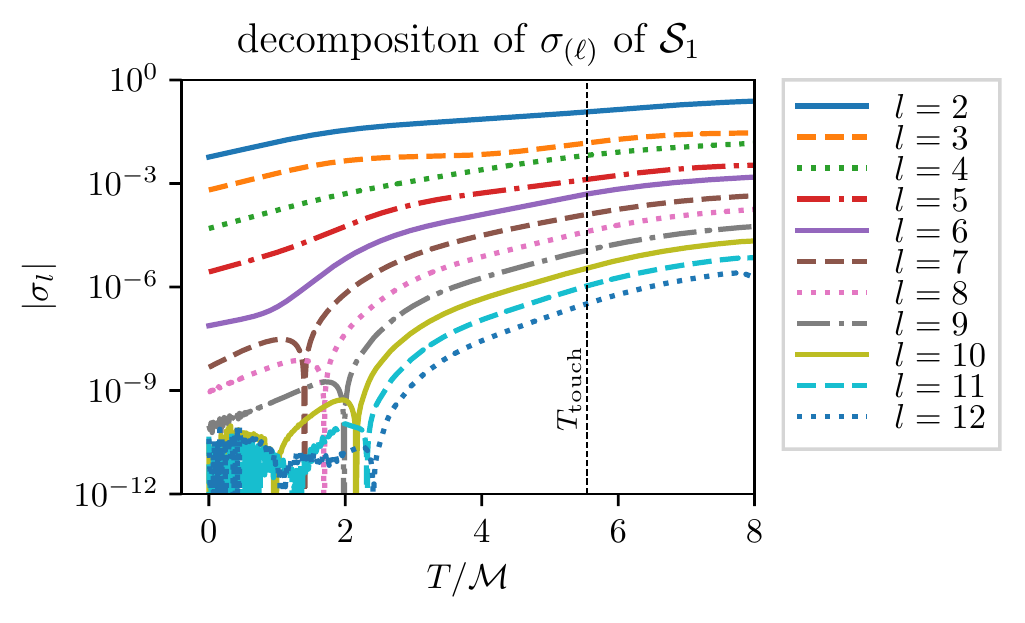}
  \includegraphics[width=\columnwidth]{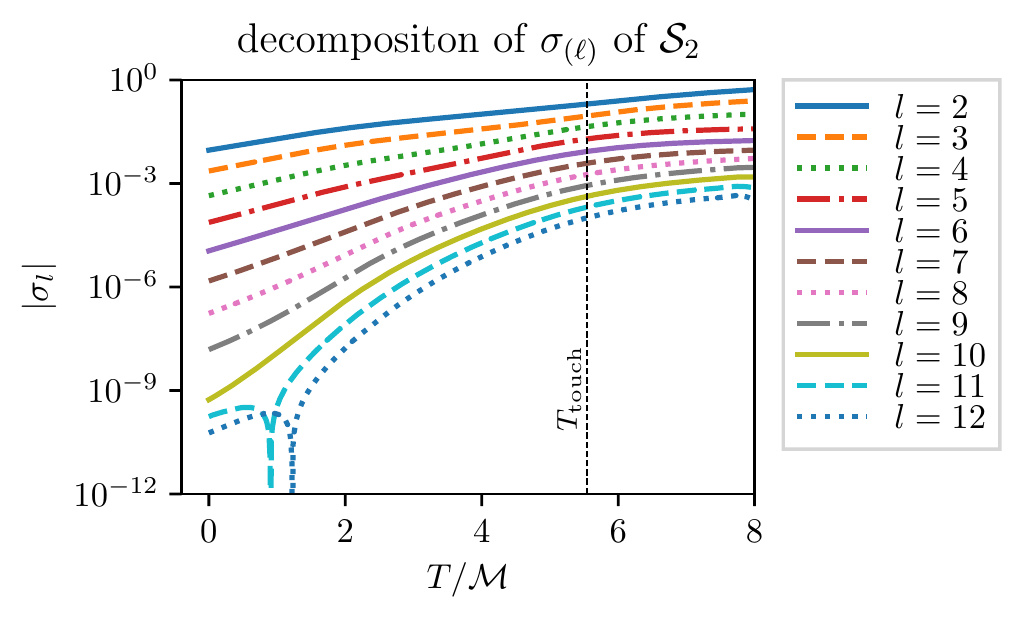}
  \caption{The mode decomposition of the shear for the two individual
    black holes.  See text for details. }
  \label{fig:shear12}
\end{figure*}

Fig.~\ref{fig:shear12} shows $|\sigma_{l}|$ for the two individual
horizons, for $l=2,3,\ldots,12$.  As the figure shows, the mode
amplitudes decrease monotonically as the mode index $l$ increases,
so that the $l=2$ mode dominates.  Similarly, as expected, the
shear generally increases with time, indicating larger fluxes as we
approach the merger.  This is confirmed by the integrals of
$|\sigma|^2$ over $\Sone$ and $\Stwo$ shown in
Fig.~\ref{fig:shear-integral}.
\begin{figure*}
  \centering    
  \includegraphics[width=\columnwidth]{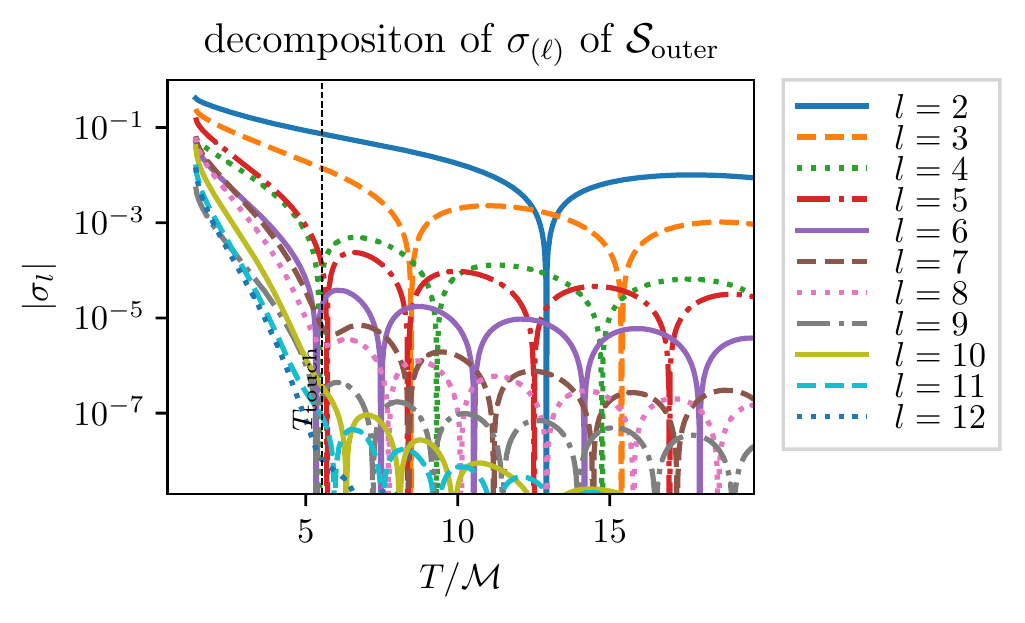}
  \includegraphics[width=\columnwidth]{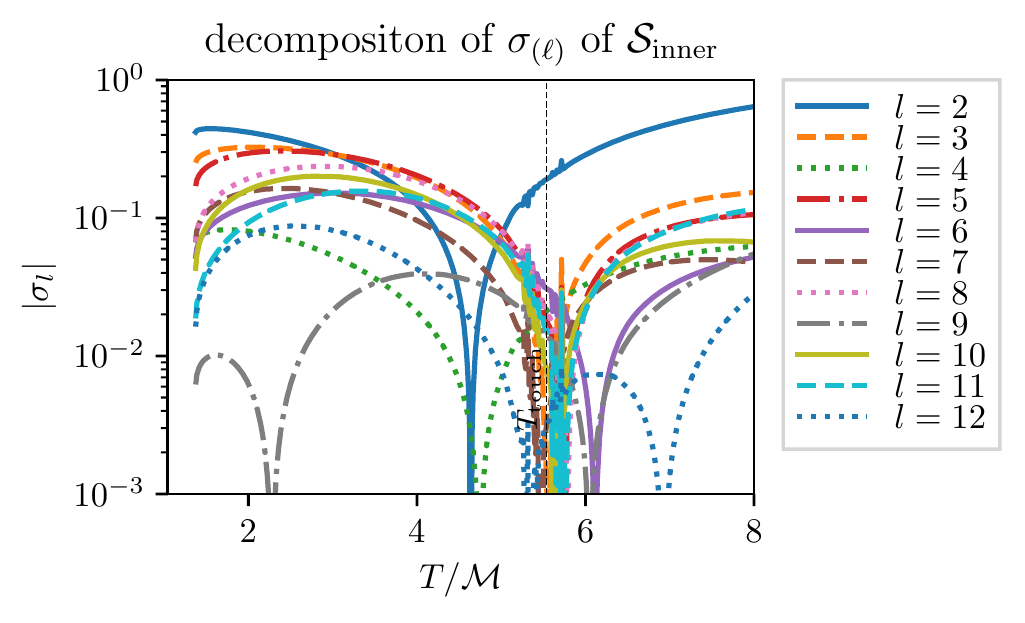}
  \caption{Shear modes for the common horizons.  The left panel shows
    $|\sigma_l|$ for the outer common horizon and the right panel
    shows the mode coefficients for the inner horizon. See text for
    further discussion.}
  \label{fig:shear-common}
\end{figure*}

\begin{figure*}
  \centering
  \includegraphics[width=0.3\linewidth]{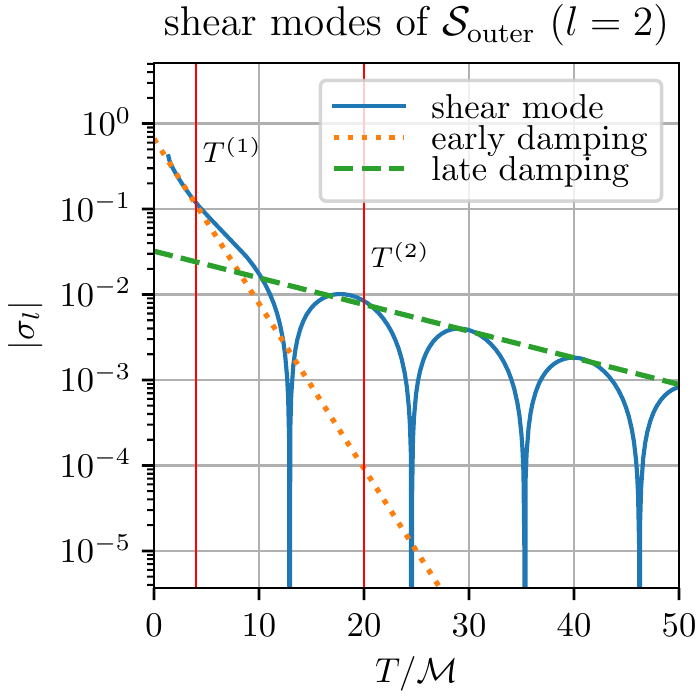}%
  \includegraphics[width=0.3\linewidth]{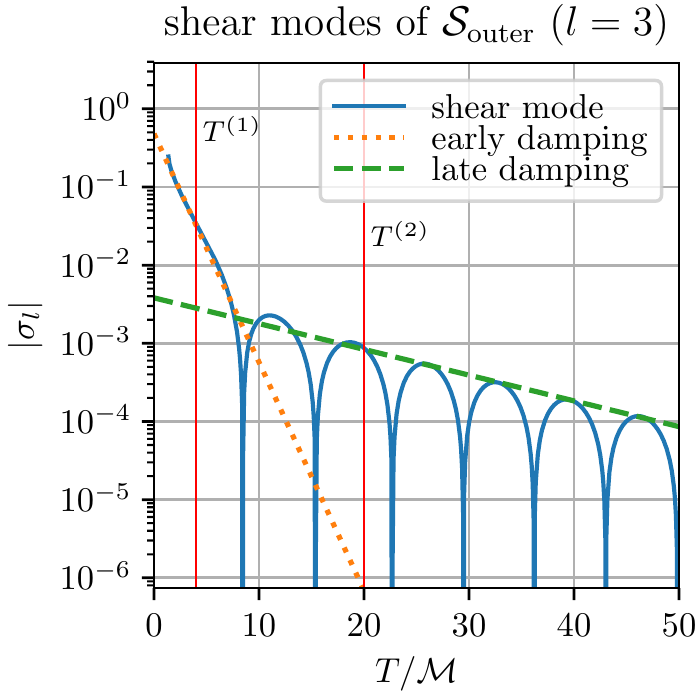}%
  \includegraphics[width=0.3\linewidth]{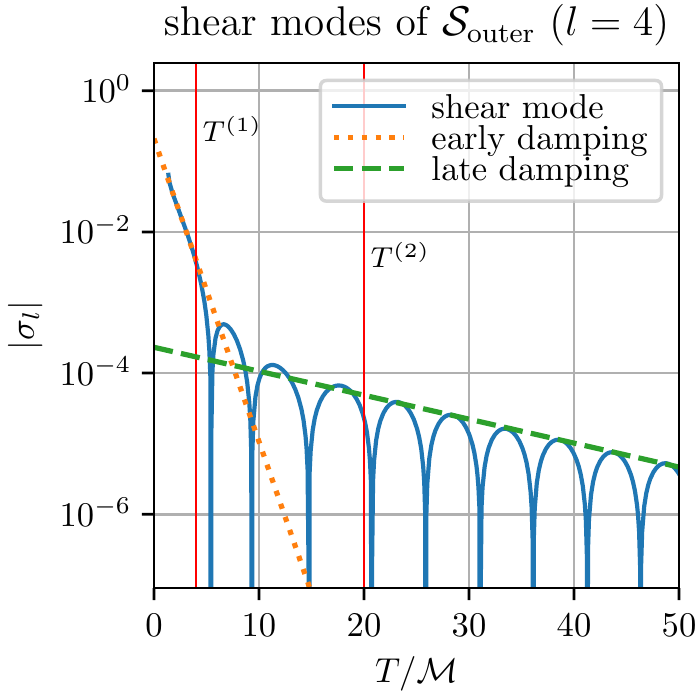}\\
  \includegraphics[width=0.3\linewidth]{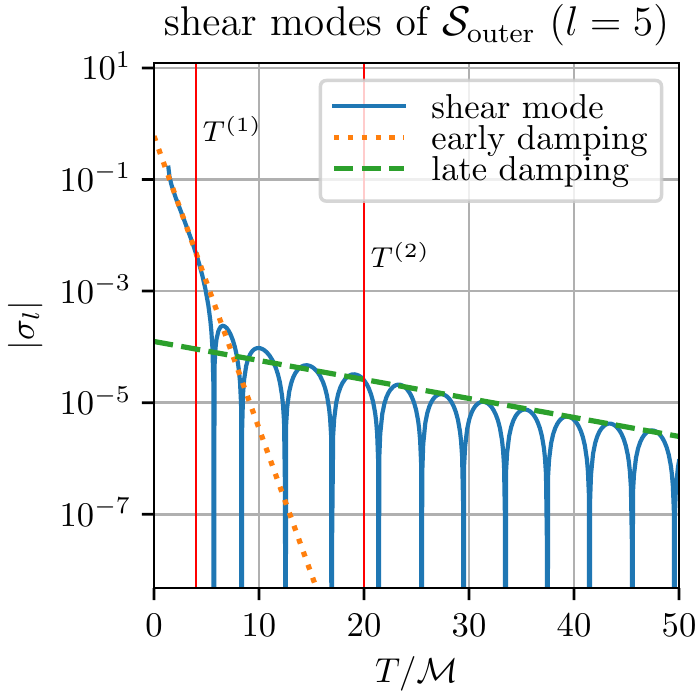}%
  \includegraphics[width=0.3\linewidth]{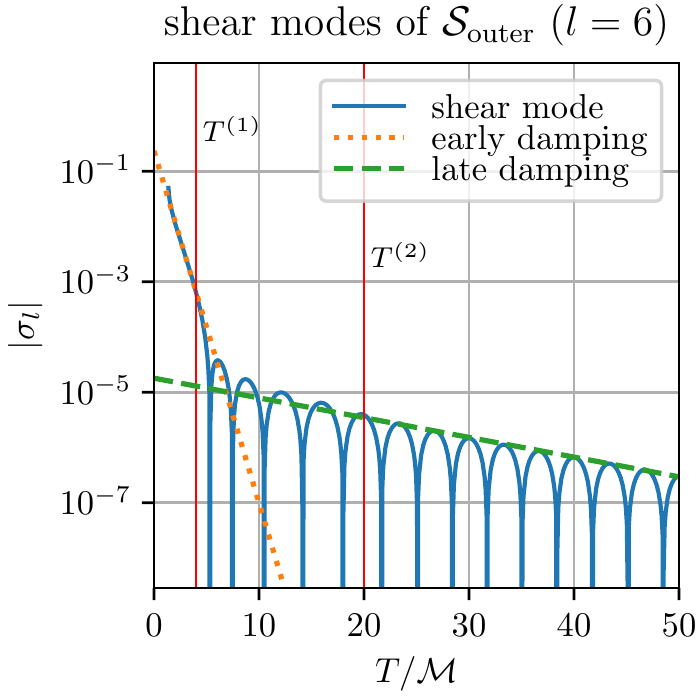}%
  \includegraphics[width=0.3\linewidth]{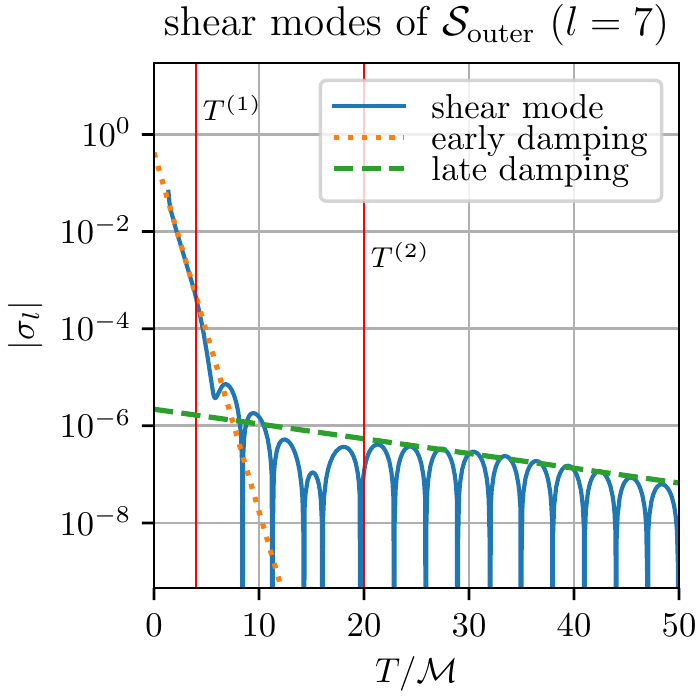}
  \caption{Fits of the shear mode for $l=2,3,\ldots,7$. The curves
    in blue show the shear amplitude, the orange dotted line shows the
    exponential fit at early times (before $T^{(1)} = 4\,\MM$), and the
    dashed green line is the exponential fit at late times (after
    $T^{(2)} = 20\,\MM$).  In each case we see a clear transition from steep
    decay to a slower decay rate. }
  \label{fig:shear-fits}
\end{figure*}

Fig.~\ref{fig:shear-common} shows the shear modes for the inner and
outer horizons.  These have a number of interesting features worth
pointing out.  Consider first the shear on the apparent horizon which
is expected to be correlated with the post-merger gravitational
waveform measured in the wavezone far away from the source.  It was
observed in \cite{Gupta:2018znn} that the horizon multipole moments
(which will be discussed below) fall-off exponentially with decay
rates consistent with the quasi-normal mode frequencies of the final
black hole. Moreover, it was shown that the fall-off of the multipole
moments is well explained by the presence of two exponentially damped
modes.  This is consistent with \cite{Giesler:2019uxc} which observed
that the post-merger waveform is well explained by the
quasi-normal modes, including the higher overtones.  Motivated by
these results, we consider a model for the shear amplitude
$|\sigma_l(t)|$ with two exponentially damped modes:
\begin{equation}
  \label{eq:two-mode-model}
  \sigma_l(t) =  A_l^{(1)} e^{\alpha_l^{(1)} t} + A_l^{(2)} e^{-i\alpha_l^{(2)} t}\,.
\end{equation}
Here we take $\alpha_l^{(1)}$ to be real, and $\alpha_l^{(2)}$
to be complex because, as shown below, at early times the shear does
not show any oscillations, while at later times it exhibits damped
oscillations.  When one mode falls off much more rapidly than the
other, a simplified piecewise-exponential model can be used:
\begin{eqnarray}
  \label{eq:two-mode-model-simple}
  \sigma_l(t) &=&  A_l^{(1)} e^{\alpha_l^{(1)} t} \,,\quad 0 < t < t^{(1)} \,,\\
  \label{eq:two-mode-model-simple2}
  \sigma_l(t) &=&  A_l^{(2)} e^{-i\alpha_l^{(2)} t}\,,\quad  t > t^{(2)}\,.
\end{eqnarray}
Again, the early part is just exponentially damped, while the later
part is an exponentially damped oscillation.  We do not necessarily
choose $t^{(1)} = t^{(2)}$.  In practice, we find that one of the
modes is rapidly decaying with an initially larger amplitude, and a
second mode which is longer lived but with lower initial amplitude.
This simplified model with suitably chosen transition times
$t^{(1,2)}$ will therefore suffice for our purposes. Before presenting
the best fit values of the decay rates, it is instructive to look at
some of the fits to the individual modes in Fig.~\ref{fig:shear-fits}.
For this figure and the following fitting results, our simulation with
the lower resolution of $1/\Delta x = 60$ and
$\tname_\text{max} = 50\,\MM$
was used in order to obtain late time data for the outer horizon $\Hout$.
It is clear from these plots that the mode amplitudes have
qualitatively different fall-offs at early and late times with the
transition occurring roughly between $\tname=8\,\MM$ and
$\tname=10\,\MM$.
It is also
clear that accurate values of $\alpha_l^{(1)},\alpha_l^{(2)}$
respectively will be obtained by taking $t^{(1)}$ as small as
possible, and $t^{(2)}$ as large as possible; we take
$t^{(1)} = 4$ and 
$t^{(2)} = 20$.  Finally, the fits of the imaginary part
$\Im(\alpha_l^{(2)})$ are obtained by considering the local maxima
of $|\sigma_l|$ after $t^{(2)}$, and the
real part $\Re(\alpha_l^{(2)})$ is obtained by looking at the
zero-crossings of $\sigma_l$.

Before looking at the best fit values obtained for the parameters in
the above model, it will be useful to keep in mind the values of the
standard quasi-normal mode frequencies for a Schwarzschild black hole.
Quasi-normal modes are defined in the framework of perturbation
theory, and they are solutions which are purely outgoing at the
horizon and at infinity
\cite{Vishveshwara:1970zz,Chandrasekhar:1975zza}. This condition leads
to a discrete set of complex frequencies labeled just by the mass of
the black hole (for spinning and charged black holes, these would be
determined by the mass, spin and charge).  The complex frequencies are
labeled by three integers $(n,l,m)$: $(l, m)$ are the usual angular
quantum numbers while $n=1,2,\ldots$ is the overtone index for the
radial wave-function.  For a Schwarzschild black hole we only need to
consider $m=0$.  Some values of the imaginary part of the frequency
are shown in Table.~\ref{tab:qnm-damping}.  Similarly, it will be
useful to know the real part of the frequency of the lowest $(n=1)$
overtone for different values of $l$.  For $l=2,3,\ldots 7$ these are
given in Table~\ref{tab:qnm-real}. Detailed data files are available
at \cite{berti-webpage}, based on \cite{Berti:2009kk,Berti:2005ys}.
It is useful to note that the imaginary frequency for a given overtone
index $n$ is fairly insensitive to the value of $l$, but for a given
$l$, the higher overtones are damped more rapidly.
\begin{table}[h]
  \caption{\label{tab:qnm-damping} Some values of the imaginary
    Schwarzschild quasi-normal-mode frequencies for different $(n,l)$
    (taken from \cite{berti-webpage}).  }
  \begin{ruledtabular}
    \begin{tabular}{c|llll}
      & $n=1$ & $n=2$ & $n=3$ & $n=4$ \\
      \hline
      $l=2$ & $-0.0890$ & $-0.2739$ & $-0.4783$ & $-0.7051$ \\
      $l=3$ & $-0.0927$ & $-0.2813$ & $-0.4791$ & $-0.6903$ \\
      $l=4$ & $-0.0942$ & $-0.2843$ & $-0.4799$ & $-0.6839$ \\
      $l=5$ & $-0.0949$ & $-0.2858$ & $-0.4803$ & $-0.6786$ \\
      $l=6$ & $-0.0953$ & $-0.2866$ & $-0.4806$ & $-0.6786$ \\
      $l=7$ & $-0.0955$ & $-0.2872$ & $-0.4807$ & $-0.6773$ \\
    \end{tabular}
    \end{ruledtabular}
\end{table}
\begin{table}[h]
  \caption{\label{tab:qnm-real} Some values of the lowest overtone
    ($n=1$) of the real Schwarzschild QNM frequency for
    $l=2,3,\ldots,7$ taken from \cite{berti-webpage}.  }
  \begin{ruledtabular}
    \begin{tabular}{llllll}
      $l=2$ & $l=3$ & $l=4$ & $l=5$ & $l=6$ & $l=7$ \\
      \hline
      $0.3737$ &  $0.5994$ &  $0.8092$ &  $1.0123$ & $1.2120$ &  $1.4097$\\
    \end{tabular}
    \end{ruledtabular}
\end{table}

At late times, we fit separately for the oscillatory and damped parts.
We fit $\Im(\alpha^{(2)}_l)$ by looking at the local maxima of
$|\sigma_l|$ and fitting them to a straight line (on a logarithmic
scale), while we fit $\Re(\alpha^{(2)}_l)$ by looking at its zero
crossings.  The fits for the early part before $4\,\MM$ turn out to
depend sensitively on the time $t^{(1)}$ in
Eq.~(\ref{eq:two-mode-model-simple}).  The choice $t^{(1)} = 4$ was
made to roughly minimize these variations. Similarly, to get
accurate values we choose to use $t^{(2)} = 20$.

Let us now look at best fit values of the exponents.  The best fit
values for $\alpha^{(1)}$ and $\alpha^{(2)}$ are shown in
Tab.~\ref{tab:shear_damping_coeffs} scaled with the ADM mass set to
unity.  Comparing the best fits for the real and imaginary parts of
$\alpha^{(2)}$ with Table~\ref{tab:qnm-real} and the first column of
\ref{tab:qnm-damping}, we find consistency over all the 6 modes
considered.  This leads us to believe that at late times the shear
modes are associated with the fundamental overtone of the quasi-normal
modes.

Things are not so clear with $\alpha^{(1)}$.  Recent work has found
that in binary black hole merger waveforms, the immediate post-merger
signal is consistent with the higher overtones of the quasi-normal
modes
\cite{Giesler:2019uxc,Okounkova:2020vwu,Forteza:2020hbw,Bhagwat:2019dtm}.
It is thus tempting to think that $\alpha^{(1)}$ should be connected
with the higher overtones.  However, comparing the best fit values of
$\alpha^{(1)}$ in Table~\ref{tab:shear_damping_coeffs} with the
complex frequencies for the higher overtones given in
Table~\ref{tab:qnm-damping}, we find no compelling evidence here. It
is possible that a combination of these higher overtones could be
considered, but we shall not attempt to do so here.

\begin{table}[h]
  \caption{\label{tab:shear_damping_coeffs}
    Fits of the shear modes based on the piecewise-exponential model
    of Eqs.~\eqref{eq:two-mode-model-simple} and
    \eqref{eq:two-mode-model-simple2}.
    We show the coefficients $\alpha^{(1)}_{l}$ of early times
    ($\tname < 4\,\MM$) and $\alpha^{(2)}_{l}$ for late times
    ($\tname > 20\,\MM$).
    For $l<7$, we estimate the errors of $\Re(\alpha_l^{(2)})$ to be about
    $1\,\%$ and of $\Im(\alpha_l^{(2)})$ to be about $10\,\%$.
    All values have been scaled to correspond to a $\MADM = 1$ simulation.
  }
  \begin{ruledtabular}
    \begin{tabular}{clll}
      $l$ &
        \multicolumn{1}{c}{$\alpha^{(1)}_{l}$} &
        \multicolumn{1}{c}{$\Re(\alpha^{(2)}_{l})$} &
        \multicolumn{1}{c}{$\Im(\alpha^{(2)}_{l})$}\\
      \hline
      $2$ & $-0.578$ & $0.377$ & $-0.093$\\
      $3$ & $-0.875$ & $0.602$ & $-0.099$\\
      $4$ & $-1.284$ & $0.798$ & $-0.102$\\
      $5$ & $-1.568$ & $1.015$ & $-0.102$\\
      $6$ & $-1.906$ & $1.217$ & $-0.107$\\
      $7$ & $-2.210$ & $1.359$ & $-0.091$\\
    \end{tabular}
    \end{ruledtabular}
\end{table}

There is so far no compelling theoretical reason to assume that the
quasi-normal frequencies should be reflected at the dynamical horizon
where the horizon is still evolving. Moreover, we have not accounted
for the particular time coordinate and gauge choices made in the
numerical simulation.  Nevertheless the agreement of $\alpha^{(2)}$
with the QNM damping times can be taken as strong evidence.  It would
of course be very interesting to find the deeper reasons for why this
correspondence happens.

\begin{figure*}
  \centering
  \includegraphics[width=0.45\linewidth]{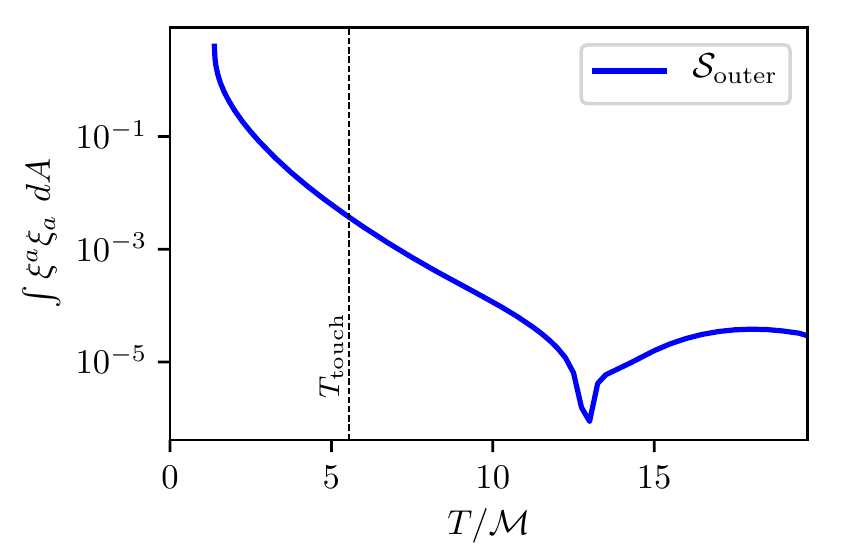}
  \includegraphics[width=0.45\linewidth]{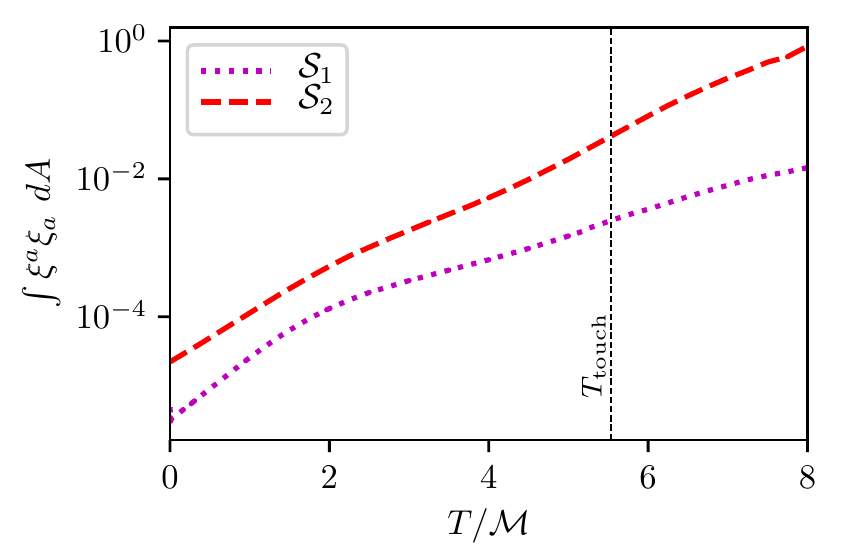}
  \caption{The behavior of the integral of $|\xi|^2$ for
    $\Hone$, $\Htwo$ and $\Hout$.  The behavior is qualitatively
    similar to $|\sigma|^2$. }
  \label{fig:xi2-integral}
\end{figure*}

\begin{figure}
  \centering
  \includegraphics[width=0.9\linewidth]{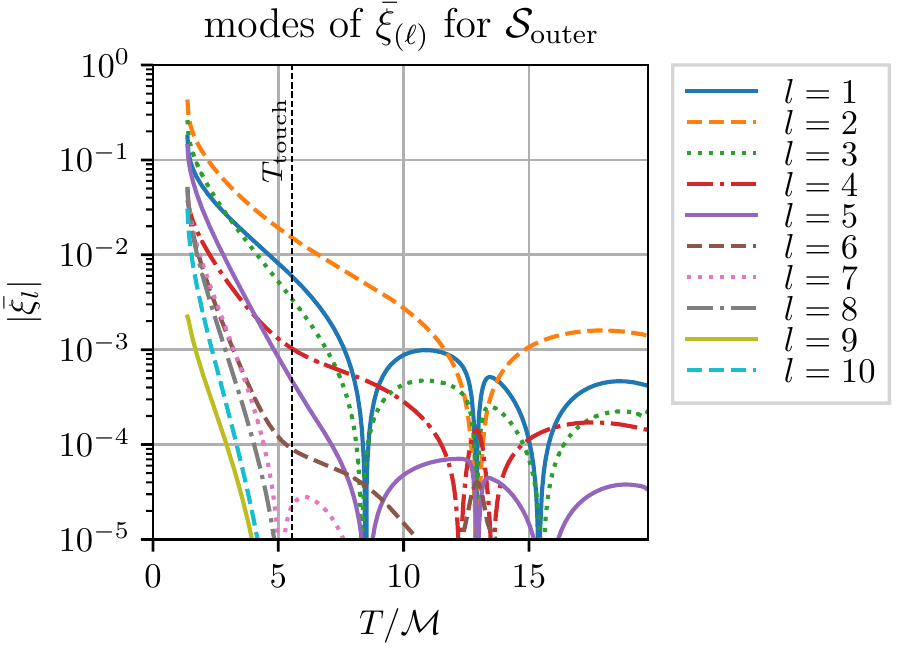}
  \caption{The mode amplitudes of $\bar{\xi}$ for $\Hout$
    for $l=1,2,\ldots 10$. Again the qualitative behavior is
    similar to the shear modes. }
  \label{fig:xi2-modes}
\end{figure}

We conclude this section by looking at the vector $\xi^a$
appearing in the flux law of Eq.~(\ref{eq:dhflux}).
Fig.~\ref{fig:xi2-integral} shows the integral of $|\xi|^2$
over the MOTS as functions of time.  The behavior is very similar to
the shear.  It turns out to be more difficult to calculate
$\xi$ numerically for $\Hin$ and we shall not do so here. Of
greater interest is the behavior for $\Hout$.  Analogous to
Eq.~(\ref{eq:shear-decomposition}), we decompose
$\bar{\xi} = \bar{m}\cdot \xi$ using spherical
harmonics of spin weight $-1$:
\begin{equation}
  \label{eq:xi-decomposition}
  \bar{\xi}(\theta,\phi,t) = \sum_{l=2}^{\infty}\sum_{m=-l}^{l}\bar{\xi}_{l m}(t){}_{-1}Y_{l m}(\theta,\phi)\,.
\end{equation}
The mode amplitudes are shown in Fig.~\ref{fig:xi2-modes}.  The
behavior is similar to the shear modes, i.e. the initial steep decay
followed by shallower decay with oscillations.  In principal, one
could attempt to compare the decay rates again, this time with the
spin-1 perturbations of Schwarzschild.  Unfortunately, we are not able
to reliably calculate the modes for longer times as we did for the
shear (the problem is $\widehat{r}^a$ at late times) and thus the best
fit values are not reliable either.  We shall not pursue this further
here.

\section{Evolution of the multipole moments}
\label{sec:moments}

\begin{figure*}
  \centering    
  \includegraphics[width=0.9\columnwidth]{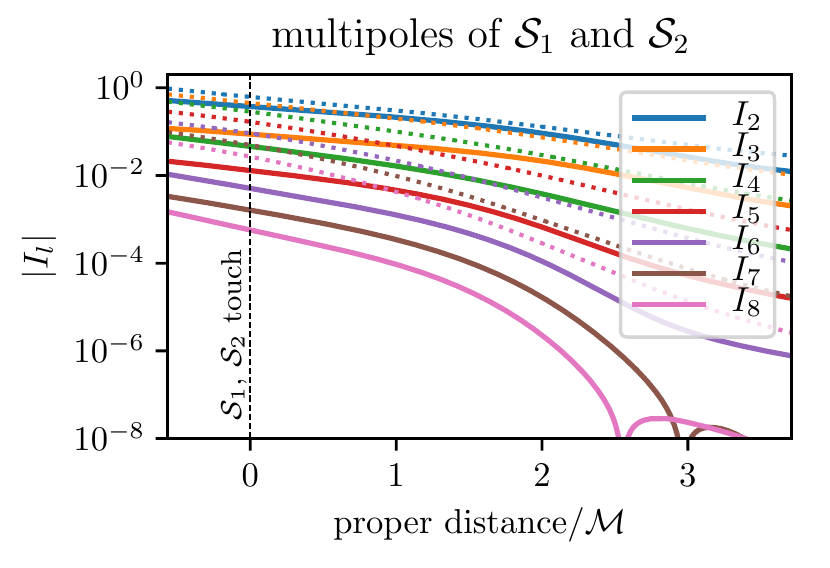}
  \includegraphics[width=0.9\columnwidth]{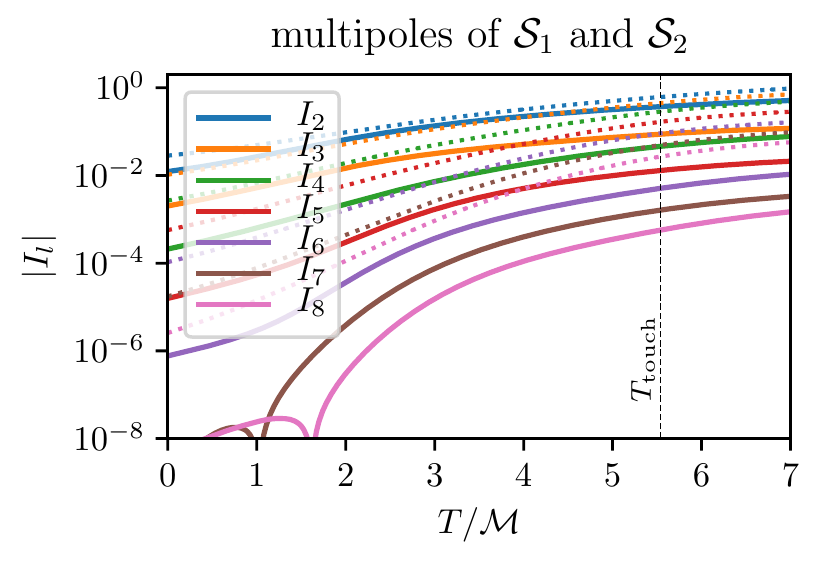}
  \caption{Multipoles of the two individual horizons as functions of
    time and separation between the black holes. In each case, the
    solid line refers to the multipoles of the smaller black hole
    $\Sone$, and the dotted line refers to the larger black hole
    $\Stwo$.  }
  \label{fig:multipoles12}
\end{figure*}

\begin{figure*}
  \centering    
  \includegraphics[width=0.9\columnwidth]{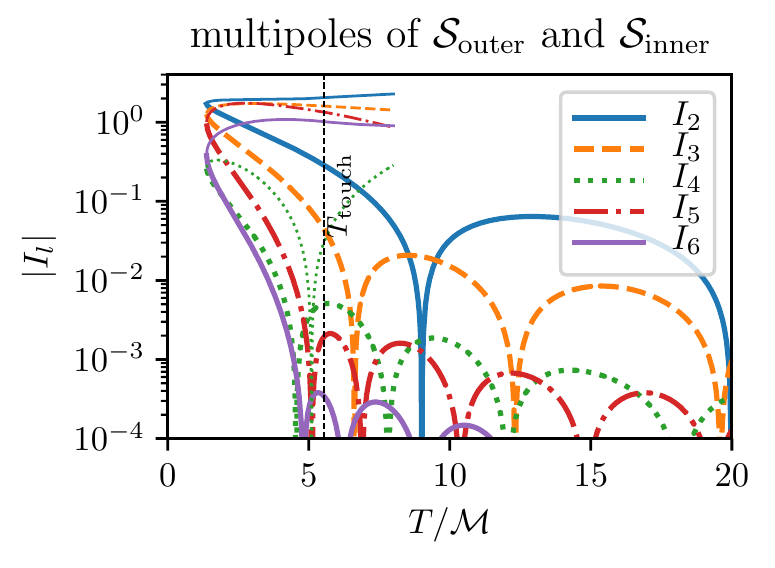}
  \includegraphics[width=0.9\columnwidth]{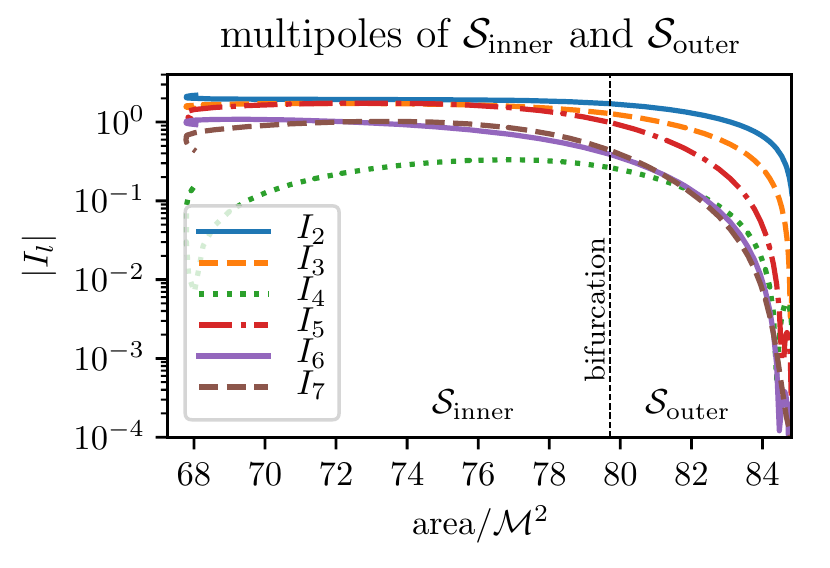}
  \caption{Multipoles of the inner and outer common horizons as
    functions of time for $l=2,3,\ldots 6$. The left panel shows the
    moments of the inner and outer horizons as functions of time.  The
    tight panel treats the inner and outer MOTSs as forming a single
    surface with the area as a coordinate.  }
  \label{fig:multipoles-common}
\end{figure*}

Turning now to the multipole moments, before we look at any results
and plots, it is clear what we should expect.  The situation is very
similar to what we have seen for the shear and stability spectrum.
First, for the individual horizons $\Sone$ and $\Stwo$, we expect at
early times to be close to Schwarzschild, i.e. all the higher moments
beyond the mass will be small.  These will increase as we get closer
to the merger, consistent with the increasing energy flux we have
encountered in the previous section.  Similarly, the common apparent
horizon $\Sout$ should show the opposite behavior, namely large higher
moments when it is formed, and settling down to Schwarzschild at later
times.  The inner horizon $\Sin$ as usual is expected to be more
complex, especially near $\ttouch$.  These expectations are borne out
in Figs.~\ref{fig:multipoles12}, and \ref{fig:multipoles-common}.  In
Fig.~\ref{fig:multipoles12}, the individual multipoles are shown both
as functions of time, and also as functions of the proper distance
between $\Sone$ and $\Stwo$.  For the common horizons, the plot as a
function of time shows the usual bifurcation between the inner and
outer horizons.  It should be kept in mind that in some sense the
distinction between $\Sin$ and $\Sout$ is artificial. Together they
form a common dynamical horizon, and excluding the time of anomalous
area increase, the area radius is a valid coordinate for the complete
dynamical horizon.  To emphasize this, in the second panel of
Fig.~\ref{fig:multipoles-common}, we plot the multipoles as a function
of the area. This shows that indeed nothing unusual occurs at the
bifurcation.  At late times, when the rate of area increase is very
small, a small increase in area represents a large duration of time
(we could have taken other quantities, for example, the shear to be a
function of area in the previous section). Similarly, we observe that
there is no unusual behavior at $\ttouch$ when the inner horizon
develops cusps and then self intersections.

\begin{figure*}
  \centering
  \includegraphics[width=0.3\linewidth]{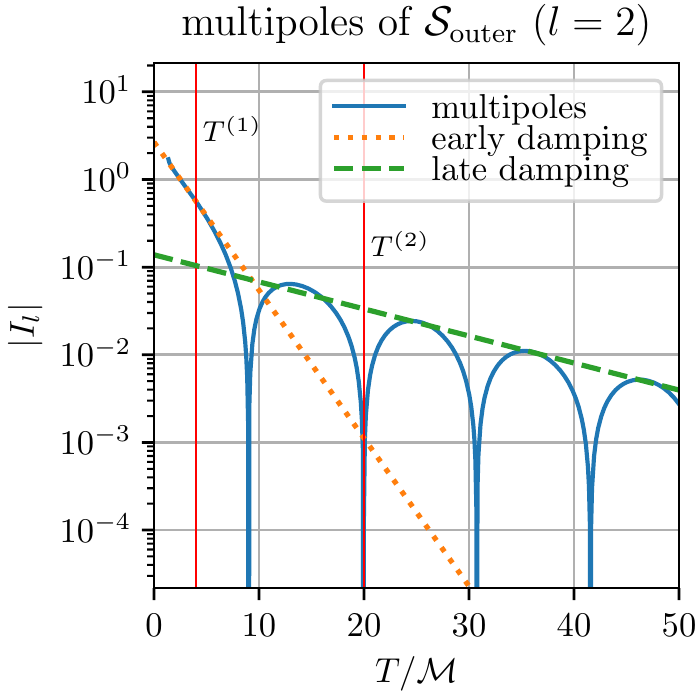}%
  \includegraphics[width=0.3\linewidth]{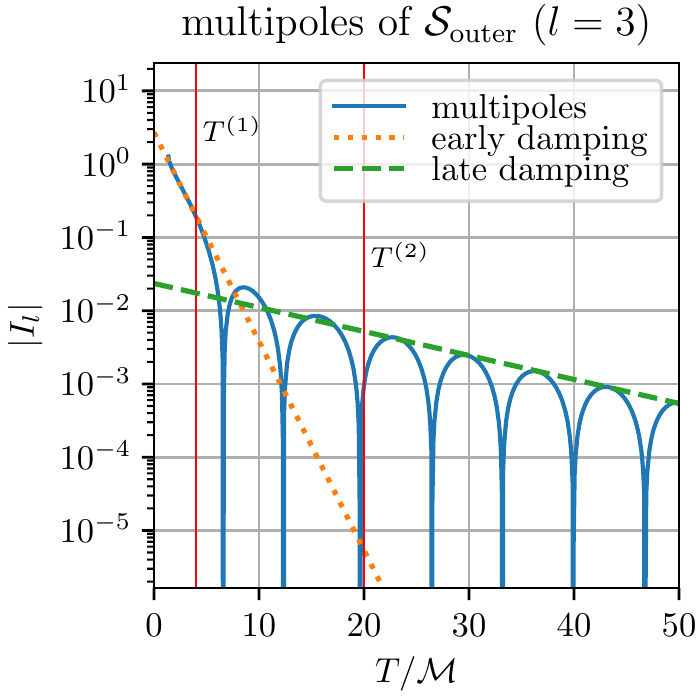}%
  \includegraphics[width=0.3\linewidth]{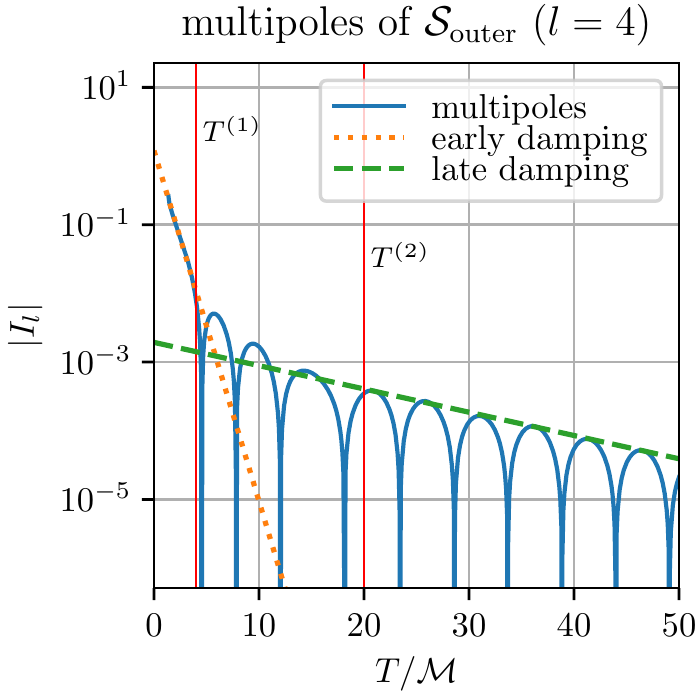}\\
  \includegraphics[width=0.3\linewidth]{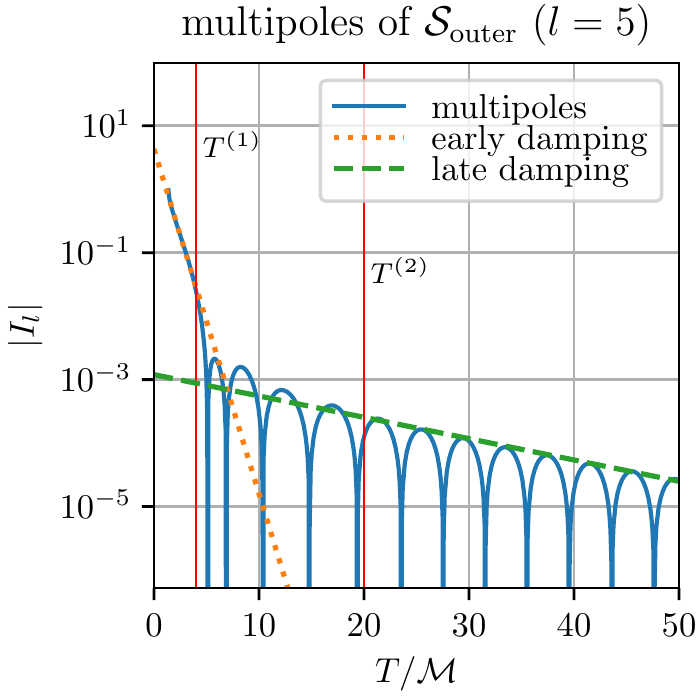}%
  \includegraphics[width=0.3\linewidth]{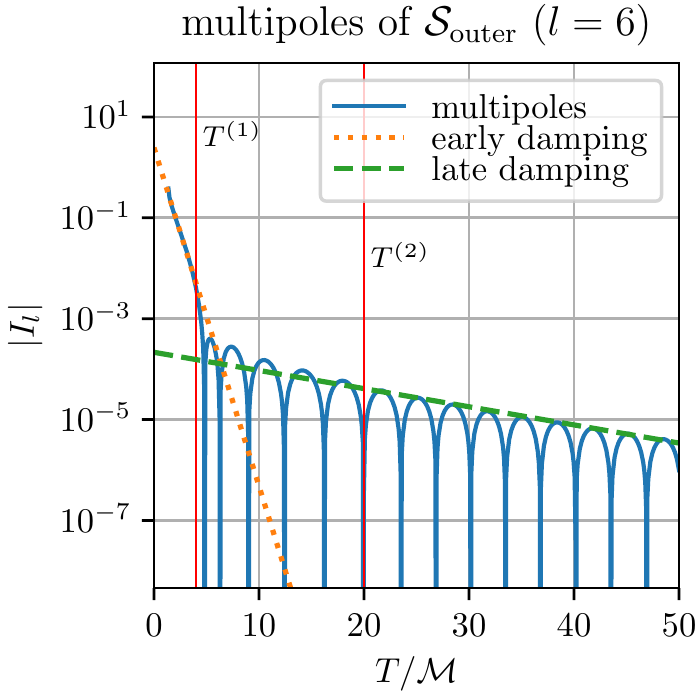}%
  \includegraphics[width=0.3\linewidth]{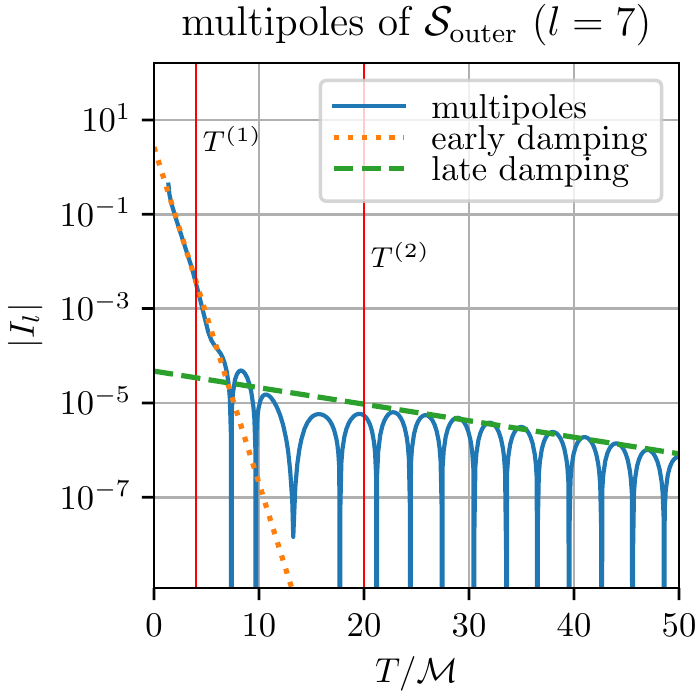}
  \caption{Fits of the multipole moments for $l=2,3,\ldots,7$. The
    plots are similar to Fig.~\ref{fig:shear-fits}. The blue curves
    are the multipole moments of the apparent horizon as functions of
    time, and fits at early and late times are also shown. }
  \label{fig:multipole-fits}
\end{figure*}

We now turn to the decay of the multipole moments for the outer
dynamical horizon.  As before, we use the two component model given in
Eq.~(\ref{eq:two-mode-model-simple}) with the early and late time
behavior analyzed, respectively, before $t^{(1)} = 4$ and after
$t^{(2)} = 20$.
The best fits
are shown graphically in
Fig.~\ref{fig:multipole-fits}.  As before, we see clear evidence for
the two regimes: a steep initial decay followed by damped oscillations.
The
best fit values are shown in
Table~\ref{tab:multipoles_damping_coeffs}.  For the late time
behavior, we again get good agreement with the fundamental quasinormal
mode frequencies and damping times. Again, the case for identifying
the early time steep decay with any of the higher overtones is not
very convincing.
\begin{table}[h]
  \caption{\label{tab:multipoles_damping_coeffs}
    Fits of the multipole moments based on
    Eqs.~\eqref{eq:two-mode-model-simple} and
    \eqref{eq:two-mode-model-simple2}
    where we chose $\tname^{(1)} = 4\,\MM$ and
    $\tname^{(2)} = 20\,\MM$.
    For $l<7$, we estimate the errors of $\Re(\alpha_l^{(2)})$ to be about
    $1\,\%$ and of $\Im(\alpha_l^{(2)})$ to be about $10\,\%$.
    All values have been scaled to correspond to a $\MADM = 1$
    simulation.
  }
  \begin{ruledtabular}
    \begin{tabular}{clll}
      $l$ &
        \multicolumn{1}{c}{$\alpha^{(1)}_{l}$} &
        \multicolumn{1}{c}{$\Re(\alpha^{(2)}_{l})$} &
        \multicolumn{1}{c}{$\Im(\alpha^{(2)}_{l})$}\\
      \hline
      $2$ & $-0.506$ & $0.377$ & $-0.092$\\
      $3$ & $-0.854$ & $0.604$ & $-0.098$\\
      $4$ & $-1.528$ & $0.796$ & $-0.101$\\
      $5$ & $-1.625$ & $1.017$ & $-0.101$\\
      $6$ & $-2.008$ & $1.222$ & $-0.108$\\
      $7$ & $-2.134$ & $1.343$ & $-0.105$\\
    \end{tabular}
    \end{ruledtabular}
\end{table}

\section{The slowness parameter}
\label{sec:slowness}

We have now seen, from apparently very different perspectives, that we
have two distinct post-merger regimes for the outer horizon $\Hout$.
The first is immediately after its formation, at $\tbifurcate$, where
we see a rapid approach to equilibrium.  Thus, the stability spectrum
becomes very close to that of a round 2-sphere, the shear modes and
multipoles decay rapidly to zero.  This regime is followed by a much
slower decay where oscillations in the various fields are easily
visible.  The decay rates and oscillations in this slower regime are
evidently associated with quasi-normal ringing.  The precise
transition appears to be a little bit before $10\,\MM$ in simulation
time.  Since $\tbifurcate \approx 1.37460\,\MM$, this corresponds to
$\approx 8\,\MM$ after the common horizon is formed.  We note here
that this time is quite consistent with observations of the waveform
extracted in the wavezone, far away from the black holes
\cite{Kamaretsos:2011um,Kamaretsos:2012bs,Bhagwat:2017tkm,Borhanian:2020ojt,Borhanian:2019kxt}.
In these works it is seen that the gravitational waveform is
consistent with the quasi-normal ringing, again beginning at about
$8$--$10\,\MADM$ after the merger (defined variously as the peak of
the luminosity or the strain amplitude).  Similarly, observational
results for the first binary black hole detection also find results
consistent with this observation \cite{TheLIGOScientific:2016src}.

In this section we would like to speculate about this transition time
from the view-point of the horizon dynamics. Is it possible to view
this at the time when the black hole transitions from the non-linear
to the linear regime?  Recent work has argued against such an
interpretation \cite{Giesler:2019uxc,Okounkova:2020vwu}. They find
that the gravitational wave signal immediately after the merger can be
described in terms of the higher overtones of the quasi-normal modes.
See also \cite{Forteza:2020hbw,Bhagwat:2019dtm}.  If this is
confirmed, then it indicates that the final black hole can be
described perturbatively immediately after its formation.  It also
makes more promising the idea of black hole spectroscopy
\cite{Dreyer:2003bv}, i.e observationally testing the black hole
no-hair theorem using the ringdown modes \cite{Isi:2019aib} (cf. also
possible caveats to this in \cite{Jaramillo:2020tuu}).

Turning now to the properties of $\Hout$, we have seen that the rapid
decay rates immediately after the merger are not consistent with any
single higher overtone.  This does not rule out the possibility that
several modes could be combined to accurately reproduce the decay
function that we observe, but we shall not attempt to do so here.
Furthermore, even if the immediate post-merger regime is
non-perturbative, it does not imply that the quasi-normal modes have
no role to play: several modes could be present and could be coupled
due to non-linear effects.  Here we wish to address this question in a
different way, namely by looking at evolution equations on $\Hout$,
identifying non-linear terms, and attempting to quantify their
importance.  We first need to identify which geometric quantities one
should consider. In principle, this question is closely tied to the
free data on $\HH$, i.e. the independent geometric fields that must be
specified on $\HH$ so that we can construct the spacetime in a
neighborhood of $\HH$.  This has been studied in
\cite{Booth:2012xm}. As expected, the extrinsic curvatures of each
MOTS in the null direction are part of this free data.  Our starting
point will be an equation we have encountered in paper I, namely the
evolution of the expansion $\Theta_{(V)}$ of the time evolution vector
$V^a$ in the membrane paradigm interpretation.  As in paper I, in
terms of the null normals from Eq.~(\ref{eq:normals}), the time
evolution vector is $V^a = b\ell^a + cn^a$, and the vector orthogonal
to $\Hout$ is $W^a = b\ell^a - cn^a$.  We define also
$\kappa^{(V)}=-n^bV^a\nabla_a\ell_b$.
\begin{figure}
  \centering
  \includegraphics[width=0.9\linewidth]{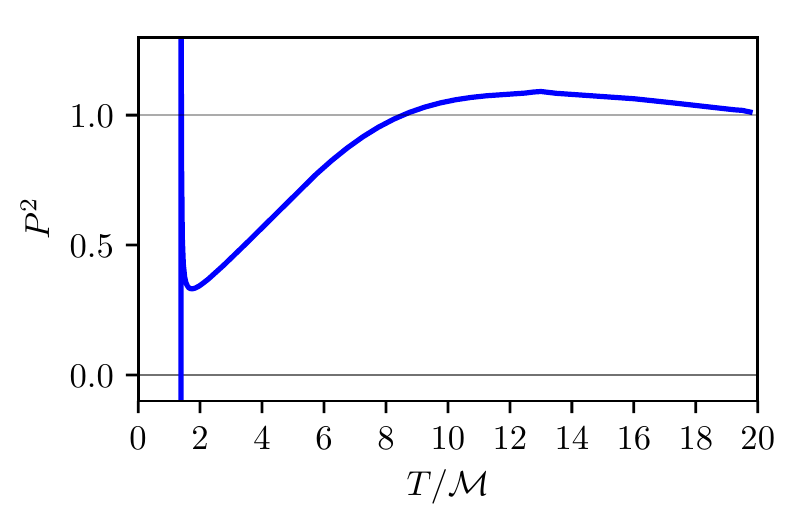}%
  \caption{The slowness parameter as a function of time.  }
  \label{fig:slowness}
\end{figure}
The qualitative average evolution of the $\HH$ can be understood in
terms of two dynamical mechanisms simultaneously in place.
Each of these  mechanisms has an associated time
scale. Following \cite{Price:2011fm} which defined
a slowness parameter using different timescales (though in a
different context), and along the lines in \cite{Jaramillo:2011rf,Jaramillo:2012rr},
we start from the equation ruling the evolution of
the expansion $\Theta_{(V)}$ encountered in Sec.~VI of paper I:
\begin{eqnarray}\label{eq:equation_theta_V_1}
  \mathcal{L}_V\Theta_{(V)} + \Theta_{(V)}^2 &=& -\kappa^{(V)} \Theta_{(V)} 
  +\sigma^{(V)}_{ab}\sigma_{(W)}^{ab} +\frac{1}{2}\Theta_{(V)}^2 \nonumber \\
  &&+ G_{ab}V^aW^b
   +  (\mathcal{L}_V\ln c) \Theta_{(V)}  \nonumber \\
  &&+  \mathcal{D}^a\left(c\mathcal{D}_ab - b\mathcal{D}_ac + 2bc\;\omega_a \right)\,.
\end{eqnarray}
Here $G_{ab}$ is the Einstein tensor.  Introducing the notion of a
``deformation rate tensor'' of ${\cal S}$ along $V^a$
(cf. e.g. \cite{Gourgoulhon:2005ng})
\begin{equation}
\Theta^{(V)}_{ab} = \frac{1}{2}{q^d}_a{q^d}_b{\cal L}_Vq_{cd} = \sigma^{(V)}_{ab} + \frac{1}{2}q_{ab}\Theta^{(V)}\, ,
\end{equation}
and analogously for $W^a$, then using $\Theta_{(W)}=-\Theta_{(V)}$, we easily get
\begin{equation}
  \Theta^{(V)}_{ab}\Theta_{(W)}^{ab} = \sigma^{(V)}_{ab}\sigma_{(W)}^{ab} -\frac{1}{2}\Theta_{(V)}^2\,.
\end{equation}
Eq.~(\ref{eq:equation_theta_V_1}) can be cast as 
\begin{eqnarray}\label{eq:evol_theta_V_2}
  \mathcal{L}_V\Theta_{(V)}   &=& -\kappa^{(V)} \Theta_{(V)} +  \Theta^{(V)}_{ab}\Theta_{(W)}^{ab} \nonumber \\
  &&+ G_{ab}V^aW^b
   +  (\mathcal{L}_V\ln c) \Theta_{(V)}  \nonumber \\
   &&+  \mathcal{D}^a\left(c\mathcal{D}_ab - b\mathcal{D}_ac + 2bc\;\omega_a \right)\,.
\end{eqnarray}
Focusing on the leading terms of the right-hand-side we identify two distinct driving mechanisms: a linear
decay term given by the $\kappa^{(V)} \Theta_{(V)}$ and a non-linear term controlled by the deformation
rate tensor of the intrinsic geometry of the surface. We expect the linear regime to be characterized
by a suppression of strong variations in the area element, and therefore a negligible value of
its ``acceleration''.  This translates into a vanishing of the left hand side in (\ref{eq:evol_theta_V_2})
as a signature of linearity. Introducing a ``decay timescale'' $\tau$ as 
\begin{eqnarray}
  \frac{1}{\tau^2} = \frac{1}{A_{\mathcal{S}}}\oint_{\mathcal S} \kappa^{(V)} \Theta_{(V)} dA\, ,
\end{eqnarray}
and an ``oscillation timescale'' $T$ controlled by the deformation rate terms
\begin{eqnarray}
  \frac{1}{T^2} &=& \frac{1}{A_{\mathcal{S}}}\oint_{\mathcal S} \Theta^{(V)}_{ab}\Theta_{(W)}^{ab}dA \nonumber \\
                  &=& \frac{1}{A_{\mathcal{S}}}\oint_{\mathcal S} \left(\sigma^{(V)}_{ab}\sigma_{(W)}^{ab} -\frac{1}{2}\Theta_{(V)}^2 \right)dA\, ,  
\end{eqnarray}
we define an instantaneous slowness parameter $P$ \cite{Price:2011fm,Jaramillo:2011rf,Jaramillo:2012rr}
as the ratio of the two  time scales
\begin{equation}\label{eq:slowness_parameter}
  P = \frac{T}{\tau}\,.
\end{equation}
Transition to the linear regime would be marked by the ``decay'' and
``oscillating'' terms becoming commensurate and therefore $P$ becoming
of order one.

Admittedly, unlike in \cite{Price:2011fm}, the identification of the
time scales with pure decay and oscillation is not so clear cut here.
We have seen that the shear also decays exponentially in time.  In any
event, regardless of this interpretation, the ratio $P$ captures the
ratio of the non-linear to linear term in
Eq.~(\ref{eq:evol_theta_V_2}).  When $P$ is close or exceeds unity,
then the non-linear term will have a correspondingly smaller
effect\footnote{It is interesting to look at $P$ from the perspective
  of the fluctuation-dissipation theorem \cite{Kubo_1966} in
  statistical mechanics. In rough terms, such a theorem states that
  (crucially, in the linear regime, near equilibrium), the relaxation
  rate and the fluctuations in a system satisfying a detailed balance
  are commensurate.  In this sense, $P\sim 1$ would mark the
  transition to a linear regime in which oscillations(/fluctuations)
  of the system equal its decay rate.  Before linearity, there is no
  reason for this relation to hold.}.  It is fairly straightforward to
calculate this quantity for $\Hout$, and the result is shown in
Fig.~\ref{fig:slowness}.  It is clear that early times after the
merger, $P$ is small indicating a larger effect of the
non-linearities, while it gets close to unity at $\approx 8\,\MM$.
The non-linear effects thus are not expected to dominate after this
time, consistent with our observations of the spectrum, shear, and
multipole moments.  Appendix~\ref{seq:Q} briefly considers the
connection between the slowness parameter developed here and the
quality factor of a resonator \cite{LalYanVyn17}; in particular
expressed in terms of quasi-normal mode frequency and damping time
\cite{Jaramillo:2011re}.

\section{Conclusions}
\label{sec:conclusions}

In this series of two papers we have studied in detail the properties
of marginally trapped surfaces in a head-on collision of two
non-spinning black holes. Even in this simple and otherwise well
studied case, we find interesting geometric and physical behavior.
Paper I has considered the status of the area increase law and the
associated geometric properties. Here in the second paper, we have
studied the stability, the time evolution of fluxes across the horizon
and the multipole moments.  We have shown that the stability spectrum
can be used to obtain greater insights into the merger process.  We
have shown that the decay of fluxes and multipole moments for the
final common horizon is consistent with the quasi-normal mode decay
time.  However, closer to $\tbifurcate$, the time when the common
horizon is formed, the decay turns out to be much steeper.  This holds
for all the modes of the shear and for the various multipole moments
as well.  The consistency with the quasi-normal mode decay times is
not understood from first principles, but it is consistent with the
idea of a strong correlation between fields on the horizon and the
usual gravitational waveform observed at infinity.  We have explored
two potential explanations of the faster decay just after
$\tbifurcate$. The first is the presence of higher overtones of the
fundamental quasi-normal mode, and the second in terms of the slowness
parameter.  Both of these could potentially explain the behavior.  As
far as the horizons are concerned, estimates of the decay rates of the
shear modes and multipoles favor the slowness parameter.

Future work will consider more generic initial configurations allowing
for the black holes to be spinning, and for generic orbits.  It should
be possible to extend our numerical methods for locating MOTSs to
these general situations.  This would allow us to tackle interesting
questions of interest from both astrophysical and mathematical
viewpoints.  For example, do the fluxes and multipole moments
generically decay at the rate consistent with the quasi-normal modes of
the final spinning black hole?  Is the early decay consistent with the
higher overtones and does the slowness parameter still provide a
viable explanation?  On the mathematical side, the stability operator
becomes non self-adjoint, and the question of stability and
zero-crossings of the eigenvalues become much more interesting and
complex.  This leads to deep connections with the spectral theory of
non-self adjoint operators which will be explored in forthcoming work.

\begin{acknowledgments}
  We thank Abhay Ashtekar, Ivan Booth and Lamis Al Sheikh for valuable comments and
  suggestions.  Research at Perimeter Institute is supported in part
  by the Government of Canada through the Department of Innovation,
  Science and Economic Development Canada and by the Province of
  Ontario through the Ministry of Colleges and Universities.
  We also thank the French EIPHI Graduate School (ANR-17-EURE-0002) and
  the Spanish FIS2017-86497-C2-1 project (with FEDER contribution) for support.
\end{acknowledgments}

\appendix


\section{Weyl's law for large eigenvalues}
\label{app:Weyl}

In this appendix we comment on the universality of the spectrum
asymptotics for large $\Lambda_n$.  In particular, Weyl's law
establishes that the asymptotics of the counting function $N(\Lambda)$
for the Laplacian is determined by geometric features of $\mathcal{S}$
\cite{Berger03,BaltesHilf}. Given the compactness of MOTSs, the
curvature terms are bounded and do not contribute to the leading
behavior, so that
\begin{equation}\label{eq:Weyl}
  N(\Lambda) \sim \frac{A}{4\pi}\Lambda + o(\Lambda)\, , \ \  (\Lambda \to \infty)\, .
\end{equation}
In the absence of a boundary the next term in the asymptotic expansion
is a constant depending on curvature and corners/cusps
\cite{BaltesHilf}.  Unfortunately, numerical precision does not allow
us to use this to probe the cusp of $\Sin$ at $\ttouch$. Finally,
inverting this relation (by naming $n=N(\Lambda_n)$) we get an
asymptotic behavior for $\Lambda_n$, for large $n$, namely
\begin{equation}\label{eq:large_lambda_n}
  \Lambda_n \sim \frac{4\pi}{A} n\, , \ \  (n\gg 1)\ .
\end{equation}

\section{Perturbative approach for $\Lambda_o$ of $L^{(-n)}$}
\label{sec:linear_lamnbda_o}

Let us introduce an $\epsilon$-dependent operator write $L^{(-n)}(\epsilon)$ 
\begin{equation}
L^{(-n)}(\epsilon)= -\Delta + \epsilon \mathcal{R}\, ,
\end{equation}
so that $L^{(-n)}$ in (\ref{eq:Ln_selfadjoint}) corresponds to
$L^{(-n)}(\frac{1}{2})$.  Certainly, $\epsilon=\frac{1}{2}$ is not a
small number. But we can explore, without any assumption of spherical
symmetry, the time at which $L^{(-n)}$ can be treated as a linear
perturbation of the Laplacian (shifted by constant). For this, we
consider the eigenvalue problem of the Laplacian
\begin{equation}\label{eq:Laplacian_spec}
-\Delta \phi_l = \Lambda^\Delta_l \phi_l 
\end{equation}
The Laplacian on the closed surface $\mathcal{S}$ is a self-adjoint
non-negative operator with vanishing smallest eigenvalue
$\Lambda^\Delta_o=0$, with constant eigenfunction normalized as
$\phi_o= \frac{1}{\sqrt{A}}$. Then, assuming $\epsilon$ small, we can
perturbatively calculate to the lowest order of the principal
eigenvalue of $L^{(-n)}(\epsilon)$ as
$\Lambda_o(\epsilon) = \Lambda^\Delta_o+ \epsilon \delta \Lambda_o$,
with
\begin{equation}
  \delta \Lambda_o = \langle  \phi_o | \mathcal{R} |\phi_o\rangle
  =\frac{1}{A}\int_{\mathcal{S}}\mathcal{R}dA = \frac{8\pi}{A}\, ,
\end{equation}
where use of the Gauss-Bonnet theorem on a topological sphere has been
made.  If we now push the perturbative expression (possibly beyond its
application range), to $\epsilon=\frac{1}{2}$, we get an estimation of
$\Lambda_o$ for $L^{(-n)}$ as
\begin{equation}\label{eq:perturbative_principal}
\Lambda_o \sim \frac{4\pi}{A} = \frac{1}{4\Mirr^2}\, ,
\end{equation}
recovering the expression for $l=0$ in (\ref{eq:spectrum-round}).

\section{A horizon multipoles inequality}
\label{seq:multipole_inequality}

To complement the discussion  in section \ref{sec:moments}
we comment here on an inequality involving horizon mass multipoles.
Given the eigenfunctions of the Laplacian spectral problem (\ref{eq:Laplacian_spec}),
the following result can be derived (details will be given elsewhere):

Proposition. {\em Given a MOTS ${\cal S}$, where we write
$\omega_a= z_a+\mathcal{D}_a\lambda$ with $\mathcal{D}_a z^a=0$
and  Laplacian eigenfunctions $\phi_l$ are normalized as $\langle \phi_l, \phi_k\rangle = \delta_{lk}$,
if the inequality}
\begin{eqnarray}
&&\sum_{l\neq 0}^\infty \left|\int_{\mathcal{S}}\left(\frac{1}{2}\mathcal{R} -z_az^a - G_{ab}\ell^an^b\right)
\left(\sqrt{A}\phi_l\right)dA\right| \nonumber \\
&&<  \int_{\mathcal{S}}\left(\frac{1}{2}\mathcal{R} -z_az^a - G_{ab}\ell^an^b\right) dA
\end{eqnarray}
{\em is satisfied, then  the MOTS is stable.}

\begin{figure}
  \centering
  \includegraphics[width=0.9\linewidth]{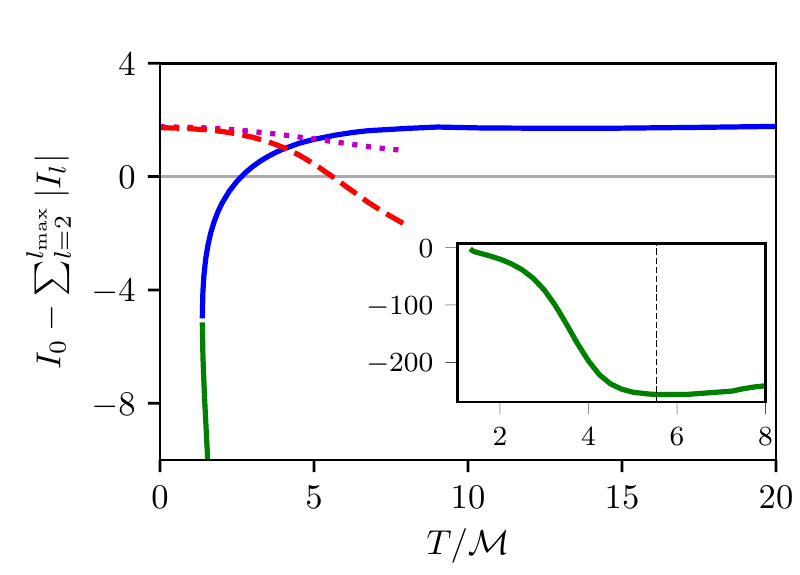}
  \caption{Multipole inequality for $\Sone$, $\Stwo$, $\Sin$ and $\Sout$. We note that
    the inequality is violated always for $\Sin$, as it should be. $\Stwo$ and $\Sout$ also
  illustrate that the inequality condition is not necessary.}
  \label{fig:multipole_inequality}
\end{figure}

In our case vacuum axisymmetric case, and using Gauss-Bonnet this reduces to
\begin{eqnarray}\label{eq:multipole_inequality}
  \sum_{l\neq 0}^\infty \left|\int_{\mathcal{S}}\mathcal{R}\phi_l dA\right|
  < \frac{8\pi}{\sqrt{A}} \ \  \Longrightarrow \hbox{MOTS stability} ,
\end{eqnarray}
that provides an inequality in terms of the horizon multipoles
introduced in \cite{Owen:2009sb}.  Therefore, if the MOTS is not
highly distorted in the sense that inequality in
(\ref{eq:multipole_inequality}) is satisfied this provides a
sufficient (but not necessary) condition for stability. Perhaps more
interestingly, its contraposition states that if the MOTS is unstable
then the inequality is violated. Even though, in a strict sense, the
result does not apply for the multipoles $I_l$ in section
\ref{sec:moments}, the fulfillment of the inequality
\begin{eqnarray}\label{eq:multipole_inequality_In}
  \sum_{l=2}^\infty \left|I_l\right|  < I_o ,
\end{eqnarray}
should provide a good estimation for a sufficient condition of MOTS stability, as indeed
confirmed in Fig.\ref{fig:multipole_inequality}.

\section{Black holes as resonators: The quality factor $Q$ and the slowness parameter}
\label{seq:Q}

If we examine the black hole resulting from a binary merger as a
resonator emitting as it is damped towards stationarity, the notion of
the slowness parameter $P$ that we have introduced makes contact
with the concept of a ``quality factor'', or Q-factor, of the
resonator. Specifically, given a resonance frequency
$\omega_n = \Omega_n - i \frac{\Gamma}{2}$, the quality factor $Q_n$
associated to $\omega_n$ is characterized as
\begin{equation}
Q_n = \frac{\Omega_n}{\Gamma}= -\frac{1}{2}\frac{\Re(\omega_n)}{\Im(\omega_n)}
\end{equation}
If we introduce, in the linear regime of the dynamics, a decay
timescale $\tau_n=1/|\Im(\omega_n)|$ and an oscillation timescale
$T_n=1/\Re(\omega_n)$ by using the QNM frequencies, then the
associated slowness parameter
$P_n = T_n/\tau_n= |\Im(\omega_n)|/\Re(\omega_n)$
(cf. \cite{Jaramillo:2011re,Jaramillo:2012rr}) is essentially the
inverse of the quality factor $P_n=\frac{1}{2Q_n}$ (this translates
into a poor Q-factor for $l=2,n=1$ of $Q\sim 2$ for Schwarzschild).
Certainly, the instantaneous $P(t)$ introduced in
(\ref{eq:slowness_parameter}) can be rescaled to match asymptotically
a value constructed from the QNM $P_n$ but, in the absence of a sound
understanding of the transition to linearity, such adjustment is just
ad hoc. More interesting in this association of a Q-factor to the
black hole is the definition of $Q_n$ in terms of the ratio of the
time-averaged energy stored in the resonator to the energy loss per
cycle
\begin{equation}
\label{e:stored_energy_power_loss}
Q_n = \Omega_n \frac{\hbox{Stored Energy}}{\hbox{Power Loss}} \ .
\end{equation}
Given that power loss can be accessed experimentally and that a value
of $Q_n$ follows from known QNMs, expression
(\ref{e:stored_energy_power_loss}) offers an avenue to assign an
energy content to the black hole, associated with a given resonant
mode. Interestingly QNM analysis has recently received much attention
in the nanoresonator optical community \cite{LalYanVyn17}, opening a
possibility for mutual transfer of tools.

\bibliography{mtt}{}

\end{document}